\documentclass[pra,superscriptaddress,twocolumn,floatfix]{revtex4}
\usepackage{amsbsy,amsmath,amsfonts,latexsym}
\usepackage{bm,color,subfigure,graphics}
\usepackage{srcltx,epsfig,float,wrapfig,psfig}
\usepackage{hyperref}
\def\bra#1{\mathinner{\langle{#1}|}} 
\def\ket#1{\mathinner{|{#1}\rangle}}

\begin{document}
\title{Super- and subradiant emission of two-level systems in the near-Dicke
  limit}      
\author{Peter G. Brooke} 
\email{pgb@ics.mq.edu.au}
\affiliation{Centre for Quantum Computer Technology and Department of Physics, 
Macquarie University, Sydney, New South Wales, Australia 2109}
\author{Karl-Peter Marzlin}
\affiliation{Institute for Quantum Information Science, 
University of Calgary, Calgary, Alberta, Canada T2N 1N4}
\author{James D. Cresser}
\affiliation{Centre for Quantum Computer Technology and Department of Physics,
Macquarie University, Sydney, New South Wales, Australia 2109}
\author{Barry C. Sanders}
\affiliation{Centre for Quantum Computer Technology and Department of Physics,
Macquarie University, Sydney, New South Wales, Australia 2109}
\affiliation{Institute for Quantum Information Science, 
University of Calgary, Calgary, Alberta, Canada T2N 1N4}
\date{\today}
\begin{abstract}
We analyze the stability of super- and subradiant states in a
system of identical two-level atoms in the near-Dicke limit, i.e.,
when the atoms are very close to each other compared to the wavelength
of resonant light. The dynamics of the system are studied using a 
renormalized master equation, both with multipolar and
minimal-coupling interaction schemes.  We show that both models lead to the
same result and, in contrast to non-renormalized models, predict that
the relative orientation of the (co-aligned) dipoles is unimportant
in the Dicke limit.  Our master equation is of relevance to any
system of dipole-coupled two-level atoms, and gives bounds on the
strength of the dipole-dipole interaction for closely spaced atoms.  
Exact calculations for small atom systems in the near-Dicke limit show the
increased emission times resulting from the evolution generated by the
strong dipole-dipole interaction.  However, for large numbers of atoms in the
near-Dicke limit, it is shown that as the number of atoms increases, the effect of the dipole-dipole interaction on collective emission is reduced.
\end{abstract} 
\maketitle
\section{Introduction}
Collective spontaneous emission from dense atomic systems has been of interest
since the pioneering work of Dicke~\cite{Dicke} 
who predicted that 
co-located two-level systems (or qubits) possess collective quantum states in
which spontaneous emission is enhanced
(superradiance) or suppressed (subradiance).
Subradiant states, which form an example of
spontaneous emission cancellation \cite{kpm07}, are of interest for quantum
information processing because they form an example of decoherence-free
subspaces (DFS) 
and subsystems~\cite{Kem01,Lidar98,Zan97a}. 
DFSs can be used to encode against system-environment
  interactions that can cause loss of quantum information. 
In the Dicke model, the system is formed by
a collection of two-level systems coupled to the vacuum radiation field.  
To achieve infinite lifetime quantum information storage, the
Dicke limit of 
co-located two-level atoms is necessary~\cite{kar07}, but
for practical implementations of quantum processors a small separation is
unavoidable and the corresponding decoherence
needs to be taken into account~\cite{Brooke07}.  Also, the
  result derived in Ref.~\cite{kar07} implies that exact superradiant
  behaviour does not exist outside of co-located atomic systems---it is
  not possible to observe perfect superradiance.  So, our results have dual
  applicability to both tests of superradiance and coherent control of atomic
  systems for quantum information.

An important question for the design of a quantum processor that makes use
of subradiant states is the trade-off between spontaneous emission 
suppression on one hand and the increase of the dipole-dipole
interaction between different two-level systems on the other hand. 
This question is difficult to answer because (i) at very short distances
the details of the interaction will strongly depend on the actual
physical particles that are modelled by the two-level system, and
(ii) the energy level shifts due to the coupling between different 
two-level systems formally diverges in the Dicke limit. Point
(i) can only be addressed by performing an ab initio calculation for
a specific system---an effort that is only justified when a promising
candidate for DFS-based quantum information processing is found.
Part of the purpose of this paper is to resolve point (ii) by
  providing a renormalized theory of the interaction of closely spaced
  two-level systems.  

The master equation we derive here is applicable to any system of
  dipole-coupled qubits.  There have been a number of examples of quantum
  processors that exploit a dipole-dipole interaction,
  e.g.,~\cite{Bei99, Br00,Pet02,Brooke07,Kif07}, all of which rely on the
  formally 
  divergent result derived in Refs.~\cite{Bela69,Lehm70i,Lehm70ii,Arg70}
and therefore can only be applied for sufficiently large separation
between the atoms. Our result allows the analysis of 
quantum processors to be extended to the near-Dicke limit.

In this paper, we use the regularized master equation to study the effect of
the dipole-dipole interaction in the near-Dicke limit.  Previous work
has approximated the dynamics of closely separated atoms using the (divergent)
contact interaction, and shown that dipole-dipole interactions
dramatically upset the predictions of the Dicke model~\cite{Gro,Car00}.  
Here, we present a renormalized model that is applicable in the near-Dicke
limit, and we apply the result to various atomic systems.  For three atoms, we
explicitly show the population transfer, caused 
by the dipole-dipole interaction, between states in the one-quantum subspace
in the near-Dicke limit.  For five atoms, we use quantum trajectories, and
show that in the near-Dicke limit, the waiting time distribution of the final
photon is extended because the dipole-dipole interaction causes
population transfer between the super- and subradiant Dicke states.  The
angular distribution confirms this, with less 
photons emitted along the interatomic axis when the
dipole-dipole interaction is included in the evolution. For $N$ atoms, we make
sensible approximations and find 
that---for a linear configuration---as the number of atoms increases the
collective spontaneous emission timescale begins to dominate the population
mixing timescale that is given by the strength of the dipole-dipole
interaction.    

The paper is organised as follows. In Sec.~\ref{sec:model} we describe
the model and present the regularized master equation for $N$
two-level atoms coupled to the radiation field.  In
Sec.~\ref{sec:rnmc}, we propose a value for both the transverse and
longitudinal regularization parameter, both of which are based on physical
considerations.  In Sec.~\ref{sec:neardicke} we show several results.  First,
that super- and subradiant emission properties are unaffected by the
dipole-dipole interaction in the exact Dicke limit.  Second, we demonstrate the
mixing properties for closely-spaced small atom systems.  Third, we show that
for five atoms the emission direction is not significantly affected by the
presence of a strong dipole-dipole interaction in the near-Dicke limit.
Finally, we show that, in the near-Dicke limit, as the number of atoms
increases, the detrimental effect of the dipole-dipole interaction on
superradiance is reduced.  In Appendix~\ref{sec:regmin}, we derive a
regularized expression for the dynamics of a system of $N$ 2LAs coupled to a
quantized electric field at zero temperature in the minimal-coupling
picture. Due to the popularity of the master equation in the
electric-dipole picture, in App.~\ref{sec:reged} we derive the same result
using the electric-dipole picture, showing that it does not matter which
description of the electric-field one uses. The major difference between the 
methods is the order of the divergence that needs to be regularized, and the
treatment of the static dipole-dipole interaction.  
\section{Regularized master equation for $N$ two-level atoms}
\label{sec:model}
We consider a system of $N$ two-level atoms interacting with the
free radiation field, which initially occupies the vacuum state.
The interaction with the field induces a dipole-moment in the
atoms, and these induced dipoles are allowed to interact via
photon exchange. 
For each atom, the matrix elements of the dipole operator are
given by the same vector
${\boldsymbol d}\equiv \langle e | \hat{{\boldsymbol d}} | g \rangle $.
In absence of the interaction, the atomic system's Hamiltonian
is given by
\begin{align}
\widehat{H}_{\text{s}} = \frac{\hbar\omega_0}{2} \sum_{n=1}^N
\hat{\sigma}_{nz} \; ,
\label{Hs}\end{align}
with $\omega_0\equiv ({\cal E}_e - {\cal E}_g)/\hbar$ the resonance frequency,
${\cal E}_i$ the atomic energy levels,
and $\hat{\sigma}_{nz} $ the Pauli z-matrix for the $n^\text{th}$ atom.
We assume that the latter is located at position ${\boldsymbol r}_n$
and use $\boldsymbol{r}_{nm} \equiv \boldsymbol{r}_n -\boldsymbol{r}_m$
to denote the distance vector between a pair $(n, m)$ of atoms.

For such a system of two-level atoms it is possible to derive
a Markovian master equation in which the influence of the radiation
field is described by a set of atomic decay rates, energy shifts, and 
coupling terms.
\cite{Bela69,Lehm70i,Arg70,Car00,Clem03}.  However, for closely
spaced atoms it is necessary to regularize and renormalize these parameters.
Using minimal coupling (see App.~\ref{sec:regmin}) and
electric dipole coupling (see App.~\ref{sec:reged})
we have derived a regularized master equation to
describe the dynamics of the density matrix $\rho$ of
an ensemble of atoms in the near-Dicke limit. 
The results of both calculations agree (see Sec.~\ref{sec:equivmc})
and yield
\begin{align}
\label{eq:mefmc}
 \dot{\rho} =& -\frac{\text{i}}{\hbar}[\widehat{H}_{\text{s}}, \rho] 
-\text{i} \sum_{n,m = 1}^N
  (\tilde{\Delta}^\perp_{nm} + \tilde{\Delta}^\parallel_{nm}) 
  [\hat{\sigma}_{n+} \hat{\sigma}_{m-}, \rho] \nonumber \\
&+ \sum_{n,m = 1}^N
  \tilde{\gamma}_{nm} 
  (2 \hat{\sigma}_{m-} \rho \hat{\sigma}_{n+} - \hat{\sigma}_{n+}
  \hat{\sigma}_{m-} \rho - \rho 
  \hat{\sigma}_{n+} \hat{\sigma}_{m-} )\; .
\end{align}
This equation has the same structure as the non-regularized master
equation, but the energy shifts 
$\tilde{\Delta}^\perp_{nm}$ and $\tilde{\Delta}^\parallel_{nm}$,
which describe
the dynamic (or transverse) and static (or longitudinal)
contribution to the dipole-dipole interaction,
remain finite as the distance between
the atoms goes to zero.
For $n=m$ they correspond to the Lamb shift for
two-level systems (see App.~\ref{sec:regmin}). The quantity
$\tilde{\gamma}_{nm}$ describes a collective spontaneous emission effect, 
i.e., incoherent de-excitation of the collective atomic state.

To facilitate the physical interpretation of Eq.~(\ref{eq:mefmc}), we
first state the non-regularized expressions for these parameters,
\begin{align}
\label{eq:mcurg}
\gamma_{nm} &=  \frac{3}{4}\gamma \{\alpha(\xi_{nm}) +
  \eta_{nm}  \beta(\xi_{nm}) \}, 
\\
  \Delta_{nm}^\perp =& \frac{3}{4}\gamma
  \Bigg \{(\eta_{nm} -1 ) \frac{\cos \xi_{nm}}{ \xi_{nm}} + (1 - 3\eta_{nm}) 
\nonumber \\
&\times \left(
  \frac{\sin \xi_{nm}}{\xi_{nm}^2} + \frac{\cos 
  \xi_{nm}}{\xi_{nm}^3}  \right) \Bigg \} - \Delta^\parallel_{nm} ,
  \label{eq:rtddnr}  
\\
\label{eq:deltaParUnreg}
\Delta^\parallel_{nm} &=  \frac{3\gamma (1 - 3\eta_{nm}) }{4
  \xi_{nm}^3},
\end{align}
which we denote by the same symbols as the regularized parameters
but without tilde. In the limit of small $\xi_{nm} \equiv k_0\, r_{nm}$,
the parameter $\gamma_{nm}$ approaches $\gamma /2$, where
$\gamma = k_0^3 |\boldsymbol{d}|^2/(3 \pi \varepsilon_0 \hbar)$
denotes the spontaneous emission rate of an isolated
atom in free space, with $k_0 = \omega_0 /c$ the wavenumber
of resonant light. 
The two functions
\begin{align}
\label{eq:alpha}
\alpha(x) & \equiv \frac{\cos x}{x^2} - \frac{\sin
  x}{x^3}  +  \frac{\sin x}{x},  \\
\beta(x) & \equiv 3\frac{\sin x}{x^3} - 3 \frac{\cos
  x}{x^2} -  \frac{\sin x}{x}, \label{eq:beta}
\end{align}
describe the emission pattern of a radiating dipole.
The directional dependence of the
dipole-dipole interaction is expressed through the
parameter 
$\eta_{nm} \equiv (\boldsymbol{d} \cdot \boldsymbol{r}_{nm})^2 
/(|\boldsymbol{d}|^2 r_{nm}^2)$. The longitudinal energy
shift $\Delta^\parallel_{nm}$ corresponds to the 
Coulomb interaction energy between two static point dipoles.
It is exactly cancelled by a corresponding term in the transverse
energy shift $\Delta_{nm}^\perp$ so that their sum 
(the curly parentheses in Eq.~(\ref{eq:rtddnr})
) is retarded. 

The result for the corresponding regularized quantities is somewhat
more involved, 
\begin{align} 
\label{eq:mcrg}
  \tilde{\gamma}_{nm} &=
  \frac{3}{4}\tilde{\gamma}  
  [\alpha(\xi_{nm}) + \eta_{nm}  \beta(\xi_{nm}) ], 
\\
  \tilde{\Delta}^\perp_{nm} &= 
  \frac{3}{4}\tilde{\gamma} 
  \Bigg \{ \frac{\text{e}^{-r_{nm} \Lambda_\perp}}{k_0 r_{nm}^3} \bigg(
  \frac{(1-3\eta_{nm})(1+r_{nm} \Lambda_\perp)}{\Lambda_\perp^2} \nonumber 
\\
  &+ (1-\eta_{nm} ) r_{nm}^2
  \bigg) 
  + (\eta_{nm} -1 ) \frac{\cos \xi_{nm}}{ \xi_{nm}} 
  + (1 - 3\eta_{nm}) 
\nonumber \\
  &\times \left( \frac{\sin \xi_{nm}}{\xi_{nm}^2}
  + \frac{\cos \xi_{nm}}{\xi_{nm}^3} - \frac{k_0^2 + \Lambda_\perp^2}{
  \Lambda_\perp^2 \xi_{nm}^3} \right) \Bigg \}, \label{eq:mcrd} 
\\
\label{eq:delp}
\tilde{\Delta}^\parallel_{nm} &= \frac{3 \gamma \text{e}^{-\frac{r_{nm}
      \Lambda_\parallel}{\sqrt{2}}}}{8\xi^3_{nm}}
\Bigg\{ \text{e}^{\frac{r_{nm} \Lambda_\parallel}{\sqrt{2}}} (2 - 6\eta_{nm}) +
(3\eta_{nm} -1) \nonumber \\
&\times (2 +\sqrt{2}r_{nm} \Lambda_\parallel)\cos
      \frac{r_{nm}\Lambda_\parallel}{ \sqrt{2}} 
+ r_{nm} \Lambda_\parallel (3
      \sqrt{2} \eta_{nm} \nonumber \\
&- \sqrt{2} + 2 r_{nm} \eta_{nm} \Lambda_\parallel) \sin
      \frac{r_{nm} \Lambda_\parallel}{\sqrt{2}}  
\Bigg\},\end{align}
where 
\begin{align}
  \tilde{\gamma} &= \gamma \frac{\Lambda_\perp^2}{\Lambda_\perp^2 + k_0^2},
\label{eq:gammaTilde}
\end{align}
but we will see below that in the near-Dicke limit it
admits a simple physical interpretation after renormalization.
The regularization parameters $ \Lambda_\parallel$ and $\Lambda_\perp$,
which will be fixed by a renormalization procedure (Sec.~\ref{sec:rnmc}),
describe the cut-off scales in the regularized theory: photons
with wave-numbers larger than $ \Lambda_\parallel, \Lambda_\perp$
no longer contribute to the interaction.

The regularization performed here leads to a self-consistent model
that is suitable for a qualitative description
of the emission dynamics of $N$ 2LAs in the near-Dicke limit.
Its range of validity extends
to atomic separations that are much smaller than the wavelength
of resonant light, but large enough so that the electronic clouds
of the atoms do not overlap and the atoms interact like point dipoles.

We remark that our analysis of the master equation extends only to second 
order in the interaction Hamiltonian and therefore does not include 
higher-order effects such as a van der Waals interaction.
It is impossible to give a general estimate of when these effects 
can no longer be neglected. One reason is that such interactions can be
important for one physical process, but unimportant for
another.  For instance, the van der Waals
interaction may be of relevance for 
the interaction between ground-state atoms, a state in which the atoms
do not possess a dipole moment, but may be negligible for the interaction
between atoms that are polarized by a light field. Another reason is
that the significance of interactions other than dipole-dipole depends 
heavily on the atomic species, or even atomic states, under consideration.
For example, the van der Waals interaction can induce
significant energy shifts at separations as large as a few
micrometers if the atoms are prepared in a Rydberg state~\cite{singer2005}.
So, for a full analysis of closely separated atoms
that would accurately predict experimental results for a given atomic 
species, an electronic many-body problem that takes
all energy levels (including the continuum) into account would need to be
solved. Such a calculation is a formidable task and beyond the
aim of this paper, which is to provide a qualitative model for
dipole-interacting two-level systems. 

\section{Renormalization}
\label{sec:rnmc}
The transverse and longitudinal parts of the master equation
represent the propagating and nonpropagating parts of the dipole-dipole
interaction. So, the two regularization parameters,
$\Lambda_\perp$ and $\Lambda_\parallel$, have different values.  
\subsection{Transverse}
\label{sec:tran}
In Sec.~\ref{sec:remc} we have shown that the divergence
of $\Delta^\perp_{nm}$ for small
atomic separations and the divergence
of the two-level Lamb shift $ \Delta_\text{Lamb}^{\text{(2-lev)}} $ 
share the same origin.
In both cases, a virtual photon is emitted by one atom and later re-absorbed;
in the case of  $ \Delta_\text{Lamb}^{\text{(2-lev)}} $ it is re-absorbed by
the same atom, in the case of the dynamic dipole-dipole interaction
$\Delta^\perp_{nm}$ it is re-absorbed by a different atom.
For this reason, we regularize both quantities using the same parameter
$\Lambda_\perp$, and fix its value by renormalizing the
Lamb shift of a single two-level atom.  We begin by recalling Bethe's famous
argument for the calculation of the Lamb 
shift~\cite{Bet47}.  

Using second-order perturbation theory, the shift in the
atomic level $\ket{a}$ due to the interaction Hamiltonian
(\ref{eq:hintmc}), for $N=1$ particle,
is given by
\begin{align}
\label{eq:betOne}
  \Delta \mathcal{E}_a = \frac{1}{6 \pi^2 \varepsilon_0\hbar c^3} 
  \sum_{b\neq a} 
  |\boldsymbol{d}_{ba}|^2 \omega_{ab}^2
  \int_0^\infty \frac{\text{d}\mathcal{E} \mathcal{E}}{\hbar
  \omega_{ab} - \mathcal{E}}. 
\end{align}
where $\omega_{ab} = (\mathcal{E}_a - \mathcal{E}_b)/\hbar$
and we have made use of Eq.~(\ref{pdRelation}).
This expression is infinite and requires renormalization.  
The energy 
of the free-electron due to its coupling to the field
is   
\begin{align}
  \Delta \mathcal{E}^\text{free}_a = 
  -\frac{1}{6 \pi^2 \varepsilon_0\hbar c^3} 
  \sum_{b\neq a}  | \boldsymbol{d}_{ba}|^2 \omega_{ab}^2 
  \int_0^\infty \text{d}\mathcal{E}.
\end{align}
Bethe proposed that the observed energy shift for $\ket{a}$ should be the
difference between the energy of the free-electron and the bound electron
\begin{align}
  \Delta \mathcal{E}_a^{\text{obs}} &= \Delta \mathcal{E}_a - \Delta
  \mathcal{E}^\text{free}_a, 
\nonumber \\ &= 
  \frac{1}{6 \pi^2 \varepsilon_0 c^3}  \sum_{b\neq a}  |
  \boldsymbol{d}_{ba}|^2 \omega_{ab}^3  \int_0^\infty
  \frac{\text{d}\mathcal{E} }{\hbar \omega_{ab}- \mathcal{E}}. 
\end{align}
The divergence in Eq.~\eqref{eq:betOne} has been reduced from linear to logarithmic.
Bethe proposed a 
cut-off to the integration that embodies the assumption that the main part of
the Lamb shift is due to the interaction of the electron with vacuum modes of
frequency small enough to justify a nonrelativistic approach.  Naturally, he
took this cut-off to be $mc^2$
\begin{align}
\label{eq:bet1}
  \Delta \mathcal{E}_a^{\text{obs}} 
  & \simeq \frac{1}{6 \pi^2 \varepsilon_0 c^3} \sum_{b\neq a}  
  |\boldsymbol{d}_{ba}|^2 \omega_{ba}^3 \ln \frac{m
  c^2}{\hbar|\omega_{ba}|} ,
\end{align} 
for $mc^2 > |\mathcal{E}_b - \mathcal{E}_a|$.  

At this point, our derivation differs from that performed by Bethe. 
Because our model is based on two-level systems only,  
the sum in Eq.~\eqref{eq:bet1} contains only a single term
($b=1$ for $a=0$ and vice versa).
So, in our model the Lamb-shift for a 2LA is given by
\begin{align}
\Delta \mathcal{E}_{\text{2LA}} &= \Delta \mathcal{E}_1 - \Delta \mathcal{E}_0,
\nonumber \\ 
&= \frac{\hbar\gamma}{\pi} \ln \frac{\lambda}{\lambda_c} ,
\end{align}
for $\lambda_c = 2 \pi \hbar/mc$ the Compton wavelength 
and $\lambda$ the wavelength of resonant light.  
Equating with  $ \Delta_\text{Lamb}^{\text{(2-lev)}} $ of
Eq.~\eqref{eq:innmc3} gives  
\begin{align}
\Lambda_\perp  \simeq  \frac{2 \pi}{\lambda_c}, 
\end{align}
where $\tilde{\gamma} \simeq \gamma$. Thus, $\Lambda_\perp/k_0 \approx
10^5$ and $\Lambda_\perp$ cuts off the higher frequency vacuum modes.  
\subsection{Longitudinal}
\label{sec:long}
Within the dipole approximation, 
resonant light cannot resolve the microscopic structure of the 2LAs.
The $r_{nm}^{-3}$ behaviour of the dipole-dipole interaction
\eqref{eq:rtddnr} can then be considered to apply only for $r_{nm} >
a$, where $a$ is some microscopic length.  
The cut-off parameter has units of
inverse length; we therefore estimate $\Lambda^\parallel$ to be
\begin{align}
\Lambda^\parallel = \sqrt[3]{\frac{1}{V_0}},
\end{align}
where $V_0 = \tfrac{4}{3} \pi a_0^3$ for $a_0$ the Bohr radius. 
This choice is in agreement with the estimate of Ref.~\cite{Vrie} 
which is based on a $T$-matrix approach for scattering of classical light
from point particles.
At separations smaller than $a_0$, the dipole
approximation breaks down, and the Lehmberg-Agarwal master equation is no
longer an appropriate description of the physical system.  
\subsection{Comparison between regularized and non-regularized interaction}
\label{sec:regnonregcomp}
\begin{figure}[t]
\begin{center}
\subfigure[]{\label{fig:delcomp1}
\includegraphics[width=5cm]{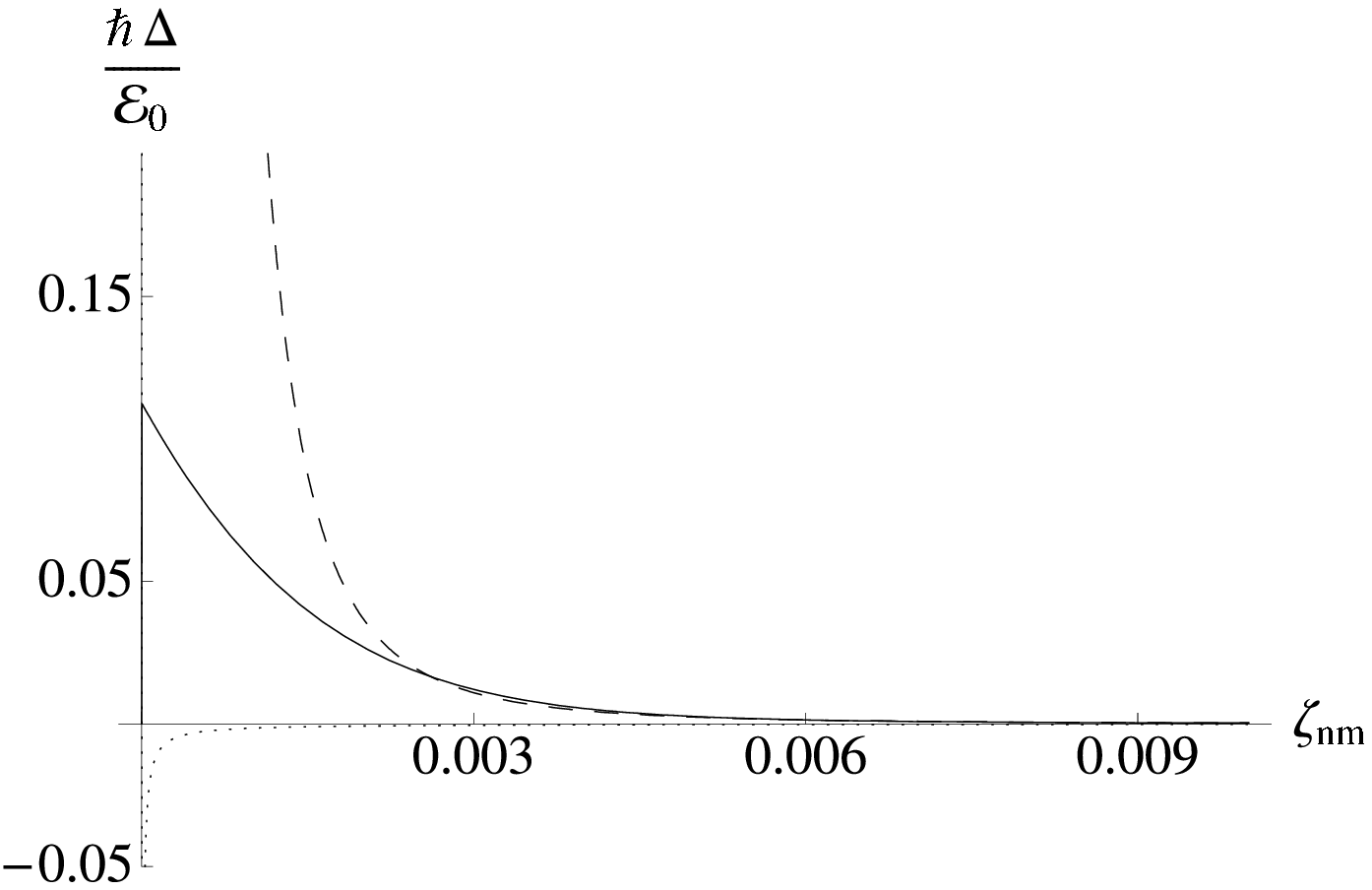}}
\subfigure[]{\label{fig:delcomp2}
\includegraphics[width=5cm]{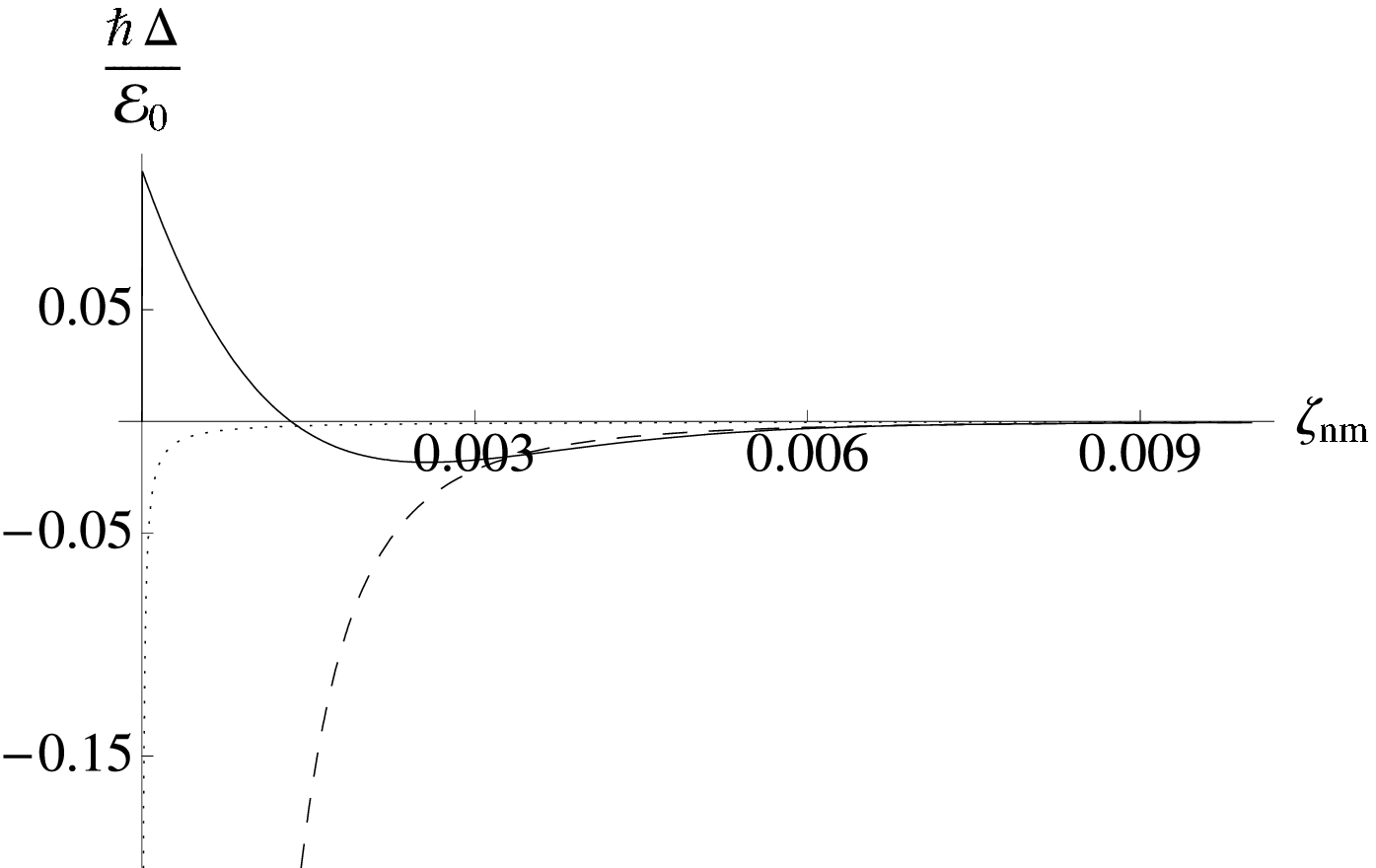}}
\end{center}
\caption{\label{fig:delcomp} 
  $\hbar \tilde{\Delta}^\perp_{nm}$ (dotted line), 
  $10^4 \hbar \tilde{\Delta}^\parallel_{nm}$ (solid line), and
  non-regularized $\Delta_{nm}^\perp + \Delta_{nm}^\parallel$ (dashed line) 
  in units of the ground state energy ${\cal E}_0$
  for (a) $\eta_{nm} = 0$ and (b) $\eta_{nm} = 1$. 
}
\end{figure}

For interatomic distances that are not substantially
smaller than the resonant wavelength $\lambda$ the predictions
of the renormalized and non-renormalized master equations agree
(see Fig.~\ref{fig:delcomp}). 
This agreement results because renormalization only affects the
description of a system at short scales. However,
in the near-Dicke limit there are substantial differences 
between the two models.

The expansion of the decay rate and energy shifts to first order
in the interatomic distance 
$\xi_{nm} = k_0\, r_{nm}$ is given by
\begin{eqnarray} 
  \tilde{\Delta}^\parallel_{nm} &=& \frac{\gamma}{4\sqrt{2}}
  \frac{\Lambda_\parallel^3}{k_0^3}
  - \frac{3\gamma(1+\eta_{nm})}{32}  \frac{\Lambda_\parallel^4}{k_0^4}
  \xi_{nm},  
\\
  \tilde{\Delta}^\perp_{nm} &=& -\frac{\gamma}{2 k_0}
  \frac{\Lambda_\perp^3}{k_0^2 +\Lambda_\perp^2 }
  + \frac{3\gamma(3-\eta_{nm})}{32}  \frac{\Lambda_\perp^2}{k_0^2} \xi_{nm}, 
\\
  \tilde{\gamma}_{nm} &=& \frac{\gamma}{2}
  \frac{\Lambda_\perp^2}{k_0^2 +\Lambda_\perp^2} \; .
\end{eqnarray} 
Hence, the decoherence part of the master equation, which is proportional to
$\tilde{\gamma}_{nm}$, takes the same form as in the exact Dicke limit.
For $\Lambda_\perp \gg k_0$, deviations of $\tilde{\gamma}_{nm}$ from the
single-atom emission rate $\gamma/2$ are only significant for larger distances
$\xi_{nm} \gtrsim 1$. 

The differences between the energy shift of the renormalized and the
non-renormalized master equation are substantial, however. 
In contrast to the non-renormalized case, the dipole-dipole interaction
does not diverge in the renormalized model. The reason for this is that
the process of regularizing a theory can be considered as ignoring
the detailed structure of a charge distribution on scales smaller
than $1/\Lambda$. Equivalently, we can think of the charge distribution
being smeared out on this scale. While the interaction energy
can diverge for point charges as they approach each other, it remains
finite for a proper (smooth, finite) charge distribution. 
Physically, atoms are composed out of many charged particles, and the
details of this charge distribution are neither captured by the
non-renormalized nor by the renormalized model. However, the
renormalized model is a more appropriate model on scales larger
than $1/\Lambda$ because it correctly describes measurable physical
quantities such as the Lamb shift (see above) or 
the static atomic polarizability \cite{Vrie}. 

Furthermore, the non-renormalized theory not only predicts 
that the energy shift diverges, but that it
diverges differently depending on the position of the
atoms relative to the orientation of their dipole moment. Hence, it would make
a difference whether the atoms approach the Dicke limit along the
direction of the dipole moment or perpendicular to it.  This is expressed
through the parameter $\eta_{nm}$ and can be seen in Fig.~\ref{fig:delcomp}.
On the other hand, the renormalized master equation predicts that the 
value of the interaction energy for $\xi_{nm} \rightarrow 0$ is
independent of the relative position of the atoms.  This is a
fundamentally different qualitative behaviour. 

To shed light on why the dipole-dipole interaction in Dicke limit
does not depend on the orientation of the (co-aligned
\footnote{In a large number of cases, 
a two-level model is insufficient to provide a consistent model
for non-aligned dipoles. The reason for this is that the dipole
moment ${\boldsymbol d}\equiv \langle e | \hat{{\boldsymbol d}} | g \rangle $
is specified by selecting a particular set of two states with
a dipole-allowed transition.  For instance, in a transition between
a $J=0$ ground state $| 1 s \rangle $
and a $J=1$ manifold of excited states $|2 p_{x,y,z} \rangle$, the
two-level system could be chosen to consist of the states
$|g \rangle = | 1 s \rangle$ and $|e \rangle = |2p_z \rangle $.
Then the dipole moment would necessarily be oriented along the $z$-axis.
In this instance, in order to include other dipole orientations
one would have to consider a four-state model.
}) dipoles, we again consider polarized atoms as 
smeared-out dipolar charge distributions.
For illustrational purposes we think of atoms as 
homogeneously polarized spheres, but any other charge distribution 
would work as well as long as it is the same for all atoms.
If the spheres are separated their interaction
energy will depend on the orientation of the distance vector
between both spheres relative to their dipole moment. 
However, when the spheres are perfectly overlapping, the distance
vector is zero and the notion of its orientation becomes meaningless.

The preceeding argument explains why there should be no
orientation dependence 
if we instantaneously place two co-aligned dipoles at the same
position.  However, the physical process of moving the dipoles to a common
position may 
induce some memory in the system (a kind of ``hysteresis''). In fact,
the non-renormalized model cannot make any sensible statement about
co-located dipoles because the energy shift is 
infinite, and it is only when we think of the two dipoles
approaching each 
other from different directions that its prediction of an orientation
dependence may appear reasonable. However, this idea is refuted
if one considers two 
polarized spheres: regardless of how the spheres are brought together,
in the Dicke limit their interaction energy should always be the same
\footnote{
Strictly speaking this argument only holds for static dipolar
charge distributions, whereas polarized atoms correspond to rotating
dipoles. We have implicitly made use of the fact that the near-field
of any oscillating charge distribution is equivalent to the static
field times an oscillating factor. Of course, when two oscillating
dipoles are moved towards the same location memory
effects could appear that are not taken into account by our simple
argument.  However, we expect that it remains valid in the limit
of infinite time, in particular if one considers energy eigenstates
as we do in Sec.~\ref{sec:neardicke}.
}.

For the particular values of the regularization parameters introduced above
we have $\Lambda_\perp ,\Lambda_\parallel \gg k_0$. If we assume that the
atomic 
dipole moment takes the value $|{\boldsymbol d}| = q a_0$, with $a_0 = 4\pi \hbar^2
\varepsilon_0 /(m q^2)$ the Bohr radius, the expansion of 
decay rate and energy shifts take the form
\begin{eqnarray} 
  \tilde{\Delta}^\parallel_{nm} &=& \frac{{\cal E}_0}{2\pi\hbar} \left (
   \frac{ 1}{\sqrt{2}}  - \frac{ 1+\eta_{nm}}{8} \frac{3^{\frac{
         4}{3}}}{2^{\frac{ 2}{3}} \pi^{\frac{1}{3}}} 
    \frac{r_{nm}}{a_0} \right ),
\label{DparExpanded} \\
  \tilde{\Delta}^\perp_{nm} &=& \frac{\gamma}{2} \frac{\lambda}{\lambda_c}
  \left ( -1 + (3-\eta_{nm}) \frac{3\pi}{8} \frac{r_{nm}}{\lambda_c} \right )
  ,  
\label{DperpExpanded}\\
  \tilde{\gamma}_{nm} &=& \frac{\tilde{\gamma}}{2} \; ,
\label{gammaExpanded} \end{eqnarray} 
with ${\cal E}_0 = m q^4/(8 h^2 \varepsilon_0^2)$ 
the modulus of the hydrogen atom's ground state energy. Thus, for our choice
of renormalization parameters $\tilde{\Delta}^\perp_{nm}$ varies on the scale
of the Compton wavelength but is much smaller than
$\tilde{\Delta}^\parallel_{nm}$, which varies on the scale of the Bohr radius.

\begin{figure}[t]
\begin{center}
\subfigure[]{\label{fig:delcomp1loglog}
\includegraphics[width=5cm]{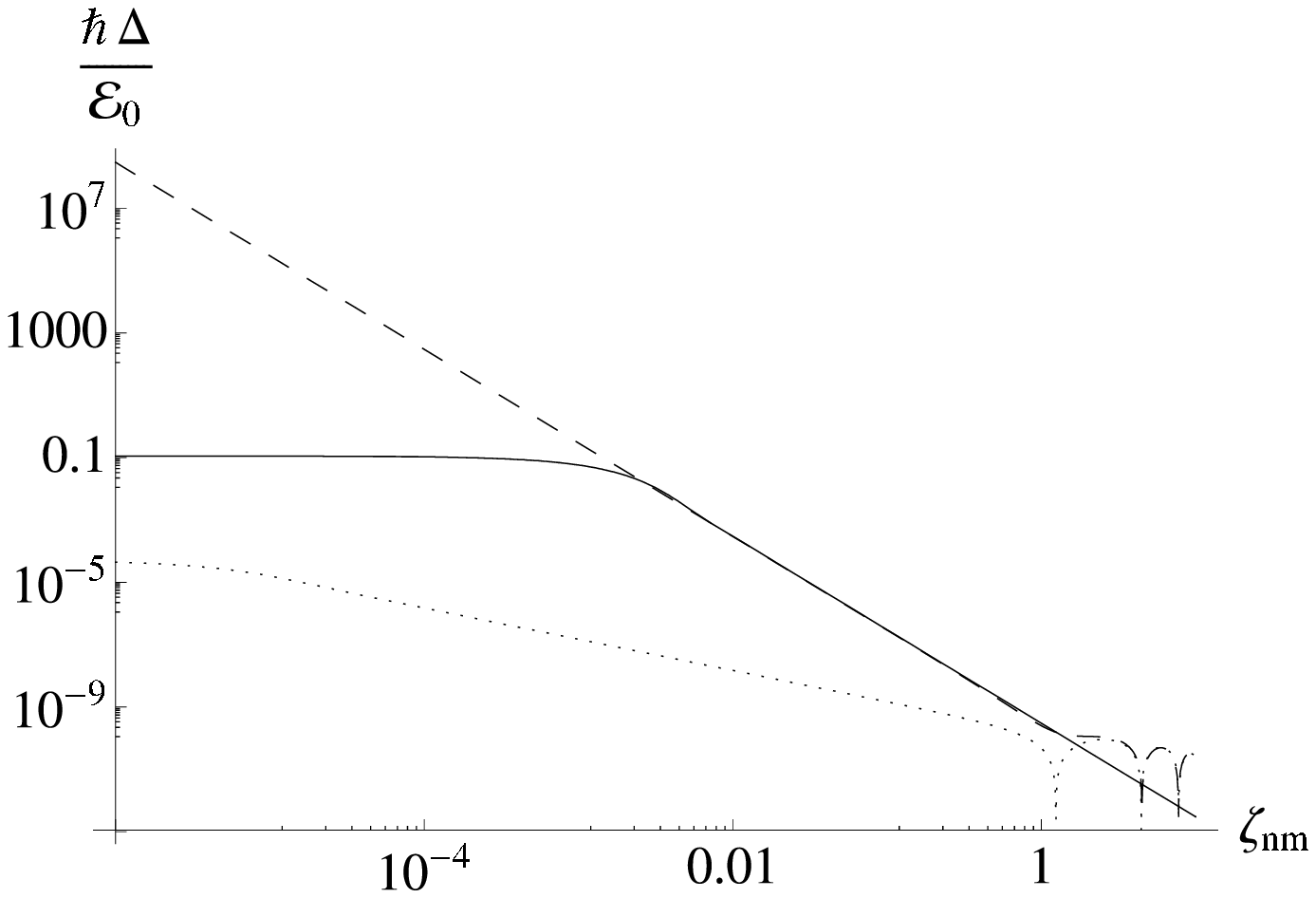}}
\subfigure[]{\label{fig:delcomp2loglog}
\includegraphics[width=5cm]{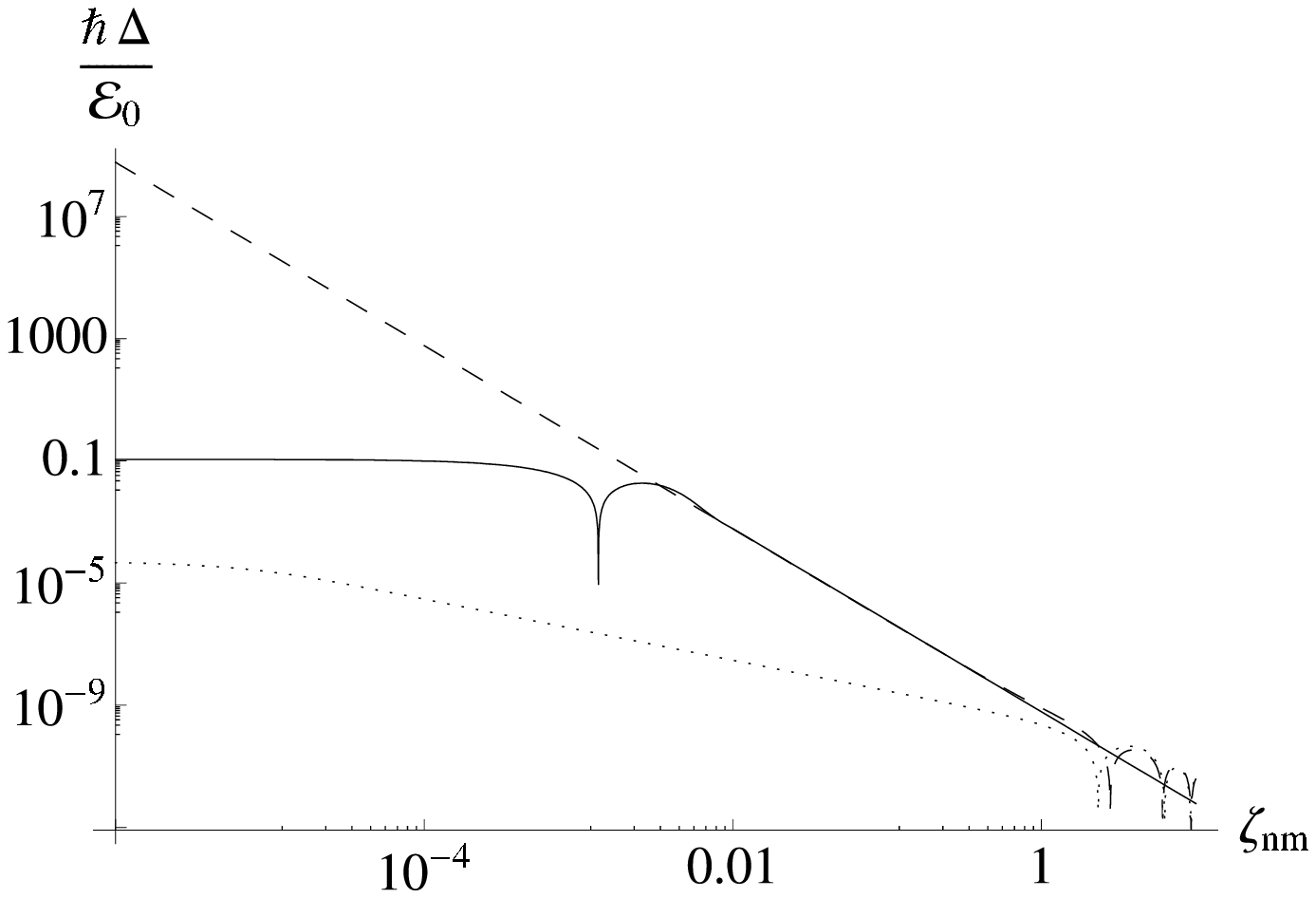}} 
\end{center}
\caption{\label{fig:delcomploglog} 
  Double-logarithmic plot of the moduli of 
  $\hbar \tilde{\Delta}^\perp_{nm}$  (dotted line), 
  $\hbar \tilde{\Delta}^\parallel_{nm}$ (solid line), and
  non-regularized $\Delta_{nm}^\perp + \Delta_{nm}^\parallel$ (dashed line) 
  in units of the ground state energy ${\cal E}_0$
  for (a) $\eta_{nm} = 0$ and (b) $\eta_{nm} = 1$.  
}
\end{figure}
Fig.~\ref{fig:delcomp} (Fig.~\ref{fig:delcomploglog}) shows the energy shifts
(modulus of the energy shifts)
as a function of the distance between two atoms. 
The sharp dips in Fig.~\ref{fig:delcomploglog} indicate
a sign change in the energy shifts. As can be seen from
Eqs.~(\ref{eq:rtddnr}) and (\ref{eq:deltaParUnreg}), the shifts oscillate
with a period of $2\pi / k_0$, which corresponds to the wavelength
of resonant light. That the dips do not approach zero is an artifact
of the double-logarithmic plot. At distances 
$\xi_{nm} \lesssim 0.5 \times 10^{-3}$,
$\tilde{\Delta}_{nm}^\perp$ begins to differ significantly from
$\Delta_{nm}^\perp + \Delta_{nm}^\parallel$. 
The region between this length scale and 
the scale $a_0$ at which the atoms start to overlap corresponds to the 
near-Dicke limit.  This is the range in which our theory 
is applicable and provides
qualitatively different predictions as compared to the non-renormalized
master equation.

Another interesting aspect of the energy shifts can be seen in 
Fig.~\ref{fig:delcomploglog}.  For distances
$\xi_{nm} > 0.5 \times 10^{-3}$, the sum
$\tilde{\Delta}_{nm}^\perp + \tilde{\Delta}_{nm}^\parallel$ agrees with
the non-renormalized dipole-dipole interaction 
$\Delta_{nm}^\perp + \Delta_{nm}^\parallel$.
To achieve this agreement for all $\xi_{nm} > 0.5 \times 10^{-3}$, 
both the transverse (or dynamic)  and the parallel (or static) energy
shift need to be taken into account. However, in the region 
$\xi_{nm}<1$ the dipole-dipole interaction is almost completely 
generated by the static energy shift, while for $\xi_{nm}>1$
it is generated by the dynamic energy shift.
This is reminiscent of the well-known fact that the near-field
of an oscillating dipole corresponds to the static dipole field
times an oscillating factor.

\section{Emission dynamics} 
\label{sec:neardicke} 
Earlier work using the divergent dipole-dipole interaction has provided a
number of insights into super- and subradiance.  In 
Ref.~\cite{Coff78} Coffey and Friedberg studied the case of two and three
atoms and showed that the dipole-dipole interaction 
causes population transfer between the super- and subradiant Dicke states.  The
authors proposed a timescale, 
$\mathcal{O}(\xi^{-3})$, for which the effects of superradiance are dampened.
In Ref.~\cite{Gro} Gross and Haroche described how the
dipole-dipole interaction generally breaks the permutation symmetry of the
atom-field couplings 
due to differences in the close-neighbor environment of the different
2LAs.  They also examined the explicit example of three 2LAs.
Outside the near-Dicke limit, Clemens {\em et al.} \cite{Clem03}  showed that
the emission of the final photon from a line of atoms showed an angular
dependence, and that this was not effected quantitatively by dipole-dipole
interactions.  

Using the results derived in the previous section, for the first time we can
investigate the effect of the dipole-dipole interaction on the
superradiance of atoms in the near-Dicke limit.  We show that in the exact
Dicke limit, in the presence of a dipole-dipole interaction, the emission is
superradiant. Next, we show the effect of the dipole-dipole interaction on the
population of the Dicke states in the near-Dicke limit for three and five 2LAs
in a linear configuration. Then, we show that in the near-Dicke limit a strong
dipole-dipole interaction does not quantitatively affect the
emission direction, but the probability of emission is reduced.  Next, we
study $N$-atom systems and find that, contrary to expectations, the denser the
atomic system (within the regime considered here), the greater the likelihood
that superradiance will dominate the emission characteristics. 

Before examining the emission dynamics in the near-Dicke limit, we show
that in the Dicke limit, 
the dipole-dipole interaction does not affect the emission dynamics.   
The bare Hamiltonian, $\widehat{H}_\text{b}$, is 
\begin{align}
\widehat{H}_\text{b} = \hbar \omega_{0} \widehat{R}_z \qquad \text{with} \qquad
\widehat{R}_{z} \equiv \sum_{n=1}^N \frac{\widehat{\sigma}_{nz}}{2} ,
\end{align}
for $N$ co-located atoms.  The renormalized dipole-dipole interaction
Hamiltonian of Eq.~\eqref{eq:mefmc} takes the form
\begin{align}
\label{eq:dipint2}
\widehat{H}_\text{d} = \sum^N_{n, m = 1} (\tilde{\Delta}^\perp_{nm} + 
\tilde{\Delta}^\parallel_{nm})
\widehat{\sigma}_{n\textrm{\tiny{+}}} \widehat{\sigma}_{m-}.
\end{align}
In the near Dicke limit we can expand this expression to first order in the
interatomic distance. Eqs.~(\ref{DparExpanded}, \ref{DperpExpanded}) yield
\begin{align} 
  \tilde{\Delta}^\perp_{nm} + \tilde{\Delta}^\parallel_{nm} &\approx
  \tilde{\Delta}_0 + \tilde{\Delta}_{1,nm} ,
\label{delta-expansion}\\
  \tilde{\Delta}_0 &\approx  \frac{{\cal E}_0}{2\sqrt{2}\pi\hbar} ,
\\
  \tilde{\Delta}_{1,nm}  &\approx
  \frac{{\cal E}_0}{2\pi\hbar} \frac{ 1+\eta_{nm}}{8} \frac{3^{\frac{
         4}{3}}}{2^{\frac{ 2}{3}} \pi^{\frac{1}{3}}} 
    \frac{r_{nm}}{a_0} . 
\end{align} 

In the Dicke limit Eq.~\eqref{eq:dipint2} can be written more succinctly by 
introducing two more collective operators~\cite{Dicke}:  
\begin{align}
\widehat{R}_{\textrm{\tiny{+}}} \equiv \sum_{n=1}^N
\widehat{\sigma}_{n\textrm{\tiny{+}}}, \quad \widehat{R}_{-} \equiv \sum_{n=1}^N
\widehat{\sigma}_{n-}, 
\end{align}
that obey $[\widehat{R}_{\textrm{\tiny{+}}},\widehat{R}_{-}]=2
\widehat{R}_{z}$ and $[\widehat{R}_{z},\widehat{R}_{\pm}]=
\pm 2 \widehat{R}_{\pm}$.  Thus, the interaction Hamiltonian becomes  
\begin{align}
\label{eq:symham}
 \widehat{H}_{\text{d}} = \widehat{H}_\text{b} + \hbar
 \tilde{\Delta}_0  \widehat{R}_{\textrm{\tiny{+}}}\widehat{R}_{-}, 
\end{align}
for $\tilde{\Delta}_{nm} = \tilde{\Delta}_0$.  
This implies that in the Dicke limit the dipole-dipole coupling strengths
become equal, independent of the distance between the atoms.  This assumption
has been used previously in deriving necessary and sufficient conditions for
  decoherence-free quantum information~\cite{Zan97a,Zan98}.  
Due to $[\widehat{H}_{\text{d}},\widehat{H}_{\text{b}}]=0$, it
is immediately seen that Dicke states are eigenstates of
$\widehat{H}_{\text{d}}$, and that the 
emission dynamics agree with those predicted by Dicke.  This occurs only for
$\tilde{\Delta}_{1,nm} = 0$, which corresponds to the exact Dicke
limit.  
\begin{figure}[t]
\begin{center}
\includegraphics[width=5cm,height=4.0cm]{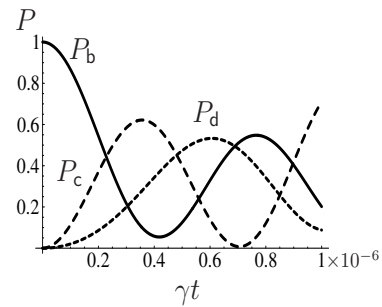}
\end{center}
\caption{ \label{fig:delevo} Populations $P_{\sf b}$, $P_{\sf c}$, $P_{\sf d}$
of the three levels {\sf b},{\sf c},{\sf d} for $\xi_{12} = \xi$, $\xi_{23} =
\xi$, and $\xi_{13} = 2\xi$ where $\xi = 0.005$, and $\eta = 1$.  The
initial state is $\ket{\sf b}$.  
}
\end{figure}
\subsection{Three atoms} 
\label{sec:3atneardicke} 
To illuminate the detrimental effect of the dipole-dipole interaction on
the emission dynamics, the special case of three atoms is considered.
For the 
purposes of this subsection, the linewidths are assumed to coincide with those
obtained in the Dicke-limit.  This is possible since at
$\mathcal{O}(r_{nm})$, $\tilde{\gamma}_{nm} = \tilde{\gamma}/2$
[Eq.~\eqref{gammaExpanded}].
For one-excitation in three 2LAs, the nonzero off-diagonal elements of
$\widehat{H}_\text{d}$ are
\begin{align}
\label{eq:dipint3at}
\widehat{H}_\text{d} =&\;  \frac{1}{\sqrt{6}}(\tilde{\Delta}_{13} 
- \tilde{\Delta}_{23})
\ket{\sf c}\bra{\sf d}  
+ \frac{2 \tilde{\Delta}_{12} - \tilde{\Delta}_{23} - \tilde{\Delta}_{13}}{3
  \sqrt{2}} 
  \ket{\sf b}\bra{\sf d} \nonumber \\
&+ \frac{\tilde{\Delta}_{23}-
    \tilde{\Delta}_{13}}{\sqrt{3}}\ket{\sf b}\bra{\sf c} + 
  \text{H.c.},   
\end{align}
for $\tilde{\Delta}_{nm} = \tilde{\Delta}_0 + \tilde{\Delta}_{1,nm} $,
and $\ket{\sf b} = 
\tfrac{1}{\sqrt{6}}(-2\ket{001} +  \ket{010} + 
\ket{100})$, $\ket{\sf c} =  \frac{1}{\sqrt{2}} (\ket{010} - \ket{100})$, and
$\ket{\sf d} = \frac{1}{\sqrt{3}} (\ket{001} +  \ket{010} + \ket{100})$. The
linewidths of $\ket{\sf b}$, $\ket{\sf c}$, and $\ket{\sf d}$ are $0$,  
$0$, and $\tfrac{3}{2} \gamma$ respectively---$\ket{\sf b}$ and $\ket{\sf c}$
are subradiant, and $\ket{\sf d}$ is superradiant.  
Setting $\tilde{\Delta}_{12} = \tilde{\Delta}_{23} = \tilde{\Delta}_{13} =
\tilde{\Delta}_0$ causes the off-diagonal terms to disappear. So, Dicke
emission dynamics are preserved for equal-strength dipole-dipole interactions
between the three 2LAs.  This can be realized by placing the 2LAs at the
vertices of an equilateral triangle.  

\begin{figure}[t]
\begin{center}
\includegraphics[width=6cm,height=4.0cm]{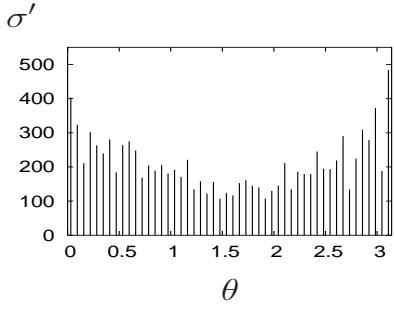}
\end{center}
\caption{Difference between polar photon counting distributions $\sigma^\prime
  = \sigma_{nd} - \sigma_{wd}$ with $( \sigma_{wd})$ and without
  $(\sigma_{nd})$ dipole-dipole interactions, for atomic orientation
   $\eta_{nm} = 0$ relative to the dipoles, and separation $r_{nm} = 1
  \times 10^{-10}$m. \label{fig:angcomp} The distributions are averaged over
  $100000$  trajectories for $0 < t < 10^4 \gamma^{-1}$.}     
\end{figure}

The master equation can be solved using the quantum trajectory method.
This unravels the evolution into a sum over records of periods of  
evolution generated by a non-Hermitian Hamiltonian that are interrupted with
stochastic jumps.  In the source-mode 
unravelling detailed in Refs.~\cite{Car00,Clem03}, the only nonzero jump
operator in the Dicke basis for $\tilde{\gamma}_{nm} = \tilde{\gamma}$ is 
\begin{align}
\hat{J} = \sqrt{\tilde{\gamma}} \sum_{n=1}^3\hat{\sigma}_{n-},
\end{align}
which means that, in the absence of a dipole-dipole interaction, the decay
cascade is $\ket{111} \to \tfrac{1}{\sqrt{3}}( \ket{110} + \ket{101} +
\ket{011}) \to \tfrac{1}{\sqrt{3}}( \ket{100} + \ket{010} +
\ket{001}) \to \ket{000}$ with probability unity.  So, in order for population
in the infinite lifetime subradiant states to 
decay, $\widehat{H}_{\text{d}}$ has to cause population transfer 
between the subradiant and superradiant states.  Fig.~\ref{fig:delevo} shows
an example of this population transfer.  Population in subradiant state
$\ket{\sf b}$ is transferred to the superradiant state $\ket{\sf d}$ on a
timescale that is fast compared to $\gamma_{\sf d}^{-1}$.    
\subsection{Five atoms} 
\label{sec:5atomsneardicke}
In Ref.~\cite{Car00} Carmichael and Kim solved the master 
equation for five atoms.  They recovered Dicke superradiance at interatomic
separations of $0.025 \lambda_0$, but with dipole-dipole interactions ignored.
They then included dipole-dipole interactions and found that the emission
dynamics returned to those expected from individual atoms.  Although five
atoms is probably too few to support a serious study of directional emission
dynamics, the photon counting records of five atoms still show a rich angular
dependence.  This implies that any influence of the dipole-dipole interaction
in the near-Dicke limit on emission direction should still be visible.   So,
in this section, the directional and temporal emission of five atoms is
examined, with the aim of better understanding the effect of the dipole-dipole
interaction in the near-Dicke limit.     
\begin{figure}[t]
\begin{center}
\includegraphics[width=6cm,height=4.0cm]{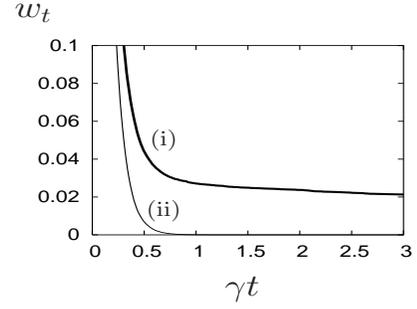}
\end{center}
\caption{Waiting time distribution of the 5th photon, $w_t =
  \bra{\bar{\Psi}(t)} 
  \widehat{B}^\dagger(t)\widehat{B}(t) \ket{\bar{\Psi}(t)} $ for 
  $\ket{\Psi(t=0)}$ the normalised state at the  time of the 4th photon
  emission, and $\ket{\bar{\Psi}(t)}$ the unnormalized state at time $t$,
  \label{fig:wtdphot5} 
  (i) including and (ii) excluding the dipole-dipole interaction for five
  atoms in a linear configuration with interatomic separation $r_{nm} = 1
  \times 10^{-10}$m.  The distributions are averaged over $10000$
  trajectories.}     
\end{figure}

The unravelling of the superradiance master equation described in
Sec.~\ref{sec:3atneardicke} accounts for emission time, but does not account
for emission direction.  In Ref.~\cite{Car00}, the authors proposed an
unravelling of the superradiance master equation that yields the
directed-detection jump operators
\begin{align}
\widehat{S}(\theta, \phi) = \sqrt{\gamma D(\theta, \phi) d \Omega} \sum_{n =
  1}^N \text{e}^{-\text{i} k_0 \vec{R} (\theta, \phi) \cdot
  \boldsymbol{r}_n} \widehat{\sigma}_{n-}, 
\end{align}   
which apply when a photon is detected in the far-field (many wavelengths
distant) within the element of solid angle $d\Omega$ in direction
$\vec{R}(\theta, \phi)$.  
The dipole radiation pattern $D(\theta, \phi) = \tfrac{3}{8
  \pi}\{ 1 - [\vec{d} \cdot \vec{R} (\theta, \phi)]^2 \} $ 
describes the emission from an isolated atom.  The
between-jump evolution is described by $\widehat{B}(t) \equiv
\text{e}^{-\text{i} \widehat{H}_B 
  t}$, where
\begin{align}
\widehat{H}_B =& \sum_{n,m = 1}^N (\tilde{\Delta}^\perp_{nm} +
\tilde{\Delta}^\parallel_{nm}) \hat{\sigma}_{n+} \hat{\sigma}_{m-} \nonumber
\\
&- \frac{\text{i}}{2} \sum^N_{n,m = 1} \tilde{\gamma}_{nm}
\hat{\sigma}_{n+} \hat{\sigma}_{m-}. 
\end{align}
Define $\delta n _\mu (\theta)$, for $\mu = 1,2, \ldots, 5$ to be the number
of times in an ensemble of quantum trajectories that photon $\mu$ is emitted
into solid angle $d\Omega$ around $\theta$, irrespective of the azimuth and
time of emission.  Then,
\begin{align}
\label{eq:sig}
\sigma(\theta) \equiv \frac{1}{d\Omega} \sum_{\mu = 1}^5 \delta n_\mu (\theta),
\end{align}
is the unconditional distribution over polar angle, summed for
photons 1 to 5. 
We calculate $\sigma(\theta)$ for five atoms in a linear configuration at 
separations of $10^{-10}$m, with and without the dipole-dipole interaction.
See Fig.~\ref{fig:angcomp}.  The cut-off implies that the region of validity of
the dipole-dipole interaction described by $\tilde{\Delta}$ is bounded below by
separations of the order of $a_0$.  In the calculations,
$\tilde{\Delta}(r_{nm} = 10^{-10} \text{m}) < \tilde{\Delta}_0$,
and $\tilde{\gamma}_{nm} \approx \tilde{\gamma}/2$.  We find that there are
  less photon emissions when the dipole-dipole interaction is included,
  implying that the dipole-dipole interaction can transfer population from
  superradiant states to 
subradiant states.  The angular distribution supports this argument, with the
difference in the number of emissions along the axis being greater than the
difference at $\theta = \pi/2$.  
\begin{figure}[t]
\begin{center}
\includegraphics[width=6cm,height=5cm]{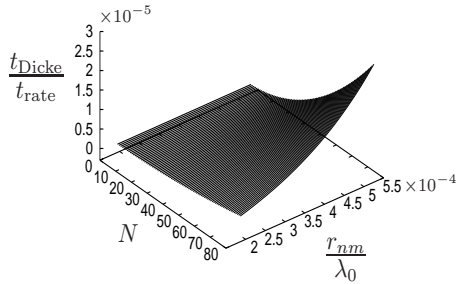}
\end{center}
\caption{ \label{fig:dipgamclose} $t_\text{Dicke}/ t_\text{rate}$
  for 
  different numbers of atoms, $N$, and separations, $r_{nm}$.  The smallest
  separation was $1.7 \times 10^{-4} \lambda_0$ and the largest $5.1 \times
  10^{-4} \lambda_0$.  For these separations,
  $\tilde{\gamma}_{nm} \gtrsim 0.99  \gamma$,
  for all $(n,m)$ validating the emission rate approximation. }  
\end{figure}

Fig.~\ref{fig:wtdphot5} shows the waiting time distribution of the
fifth photon with and without the dipole-dipole interaction.  Without
the dipole-dipole interaction, the Dicke model is recovered.  However, when the dipole-dipole interactions are included, population is transfered from subradiant states to super-radiant states, 
leading to the extended tail seen in
Fig.~\ref{fig:wtdphot5}.     
\subsection{$N$ atoms} 
\label{sec:natomsneardicke}
For many atom systems it is not possible to solve the master equation
exactly---sensible approximations are required for anything more than a few
atoms.  However, it is possible to estimate the relevant timescales for Dicke
dynamics in the near-Dicke limit.  Fig.~\ref{fig:delevo} gives some indication
of the relevant timescales.   For small times, Dicke dynamics are
preserved.  This can be quantified as follows.  The unitary evolution
of an arbitrary $N$-atom state, $\ket{\Psi}$ is
\begin{align}
\ket{\Psi(t)} = \text{e}^{-\text{i} \widehat{H}_{\text{d}} t}
\ket{\Psi(0)}, 
\end{align}
with $\widehat{H}_\text{d}  $ of Eq.~(\ref{eq:dipint2}).
So, the (shortest) timescale for Dicke dynamics in the near-Dicke limit is
\begin{align}
\label{eq:tdd}
t_{\text{Dicke}} \sim || \widehat{H}_\text{d} ||^{-1} ,
\end{align}
where $|| \cdot ||$ denotes maximum eigenvalue.  Typical values for
$ \tilde{\Delta}^\parallel_{nm}+ \tilde{\Delta}^\perp_{nm}$  in the near-Dicke
limit are $\sim 10^{8} \gamma_0$.  
Eq.~\eqref{eq:tdd} is an estimate only, and
implicitly assumes that the dipole-dipole interaction matrix and the
spontaneous emission matrix are not simultaneously diagonalizable.  Because of
the spatial dependence of $\widehat{H}_\text{d}$, this is
always true for any system consisting of more than four atoms. 

As well as estimating an upper limit on the timescale for Dicke dynamics, it
is possible to estimate the superradiant emission rate for separations ${\cal
  O}(r)$.   We neglect the ${\cal O}(r^2)$ corrections to
  $\tilde{\gamma}_{nm}$ and $ \tilde{\Delta}^\parallel_{nm}+
  \tilde{\Delta}^\perp_{nm}$.  At (finite) separations of ${\cal O}(r)$, the
  dipole-dipole interaction and collective spontaneous emission rate can be
  described by Eqs.~\eqref{DparExpanded},~\eqref{DperpExpanded}
  and~\eqref{gammaExpanded}. 
  The collective decay rates then correspond to those predicted by Dicke---the
  eigenspectrum of $(\tilde{\gamma}_{nm})$ is the same as that obtained in the
  Dicke limit.  So, we can approximate the maximum emission rate as
\begin{align}
t_{\text{rate}} = \left\{ \frac{\tilde{\gamma}}{2} N \left(\frac{1}{2} N + 1
\right) \right\}^{-1}, 
\end{align}
which occurs for $\tfrac{1}{2} N$ excited atoms in an $N$ atom system.  

To estimate the competing timescales, we take the ratio
$t_\text{Dicke} /t_{\text{rate}}$.  We assume that $t_\text{Dicke}$ is of the
order of the quickest (population) mixing rate for $N$ excited atoms, so the
ratio $t_\text{Dicke} /t_{\text{rate}}$ compares the fastest mixing timescale
(for systems of more than four atoms) with the superradiant decay timescale.  

Fig.~\ref{fig:dipgamclose} shows the effect of the interatomic
separation and the number of atoms on the emission rate.  As the number of
atoms increases, the superradiant emission rate begins to dominate over the
mixing induced by the dipole-dipole interaction.  We emphasize that
our results are only valid for atomic systems in the near-Dicke limit, i.e.,
when the spatial extent of the atomic system ensures that
Eqs.~\eqref{DparExpanded},~\eqref{DperpExpanded}, and~\eqref{gammaExpanded}
are satisfied for all atoms.
\section{Conclusion} 
\label{sec:sum} 
We renormalized the divergent superradiance master equation in
both the electric-dipole and minimal-coupling pictures.  The propagating and
nonpropagating parts of the dipole-dipole interaction were accounted for 
separately, and two values for the relevant cut-offs were proposed.  Then, it
was shown explicitly for small numbers of atoms that the dipole-dipole
interaction causes population transfer between the super- and
subradiant Dicke states, resulting in increased emission times.  This was
confirmed by the directional emission counting distribution for five atoms
that showed less 
emissions along the interatomic axis.  It was then 
shown for large numbers of atoms in the near-Dicke limit that as the number of
atoms increases, the effect of the dipole-dipole interaction on the emission
dynamics for 2LAs in a linear arrangement is reduced.  
\section{Acknowledgments}  
This project has been funded by iCORE, CIFAR, NSERC, CQCT, and Macquarie
University. One of the authors (P.G.B.) acknowledges support and hospitality during
his extended stay at IQIS at the University of Calgary. We thank 
Sir Peter Knight for discussions and Martin Kiffner for helpful 
comments on the manuscript.
\appendix
\section{Derivation of the regularized master equation
in minimal coupling}
\label{sec:regmin} 
In order to derive the master equation of the atomic system in
minimal-coupling, the matrix elements of the dipole 
operator ${\boldsymbol d}\equiv \langle e | \hat{{\boldsymbol d}} | g \rangle $
are related to the matrix elements of the electronic momentum operator
by 
\begin{equation} 
\label{pdRelation}
  {\boldsymbol p} \equiv \langle e | \hat{{\boldsymbol p}} | g \rangle 
  = \frac{\text{i}\, m}{\hbar q} ({\cal E}_e - {\cal E}_g) 
  \langle e | \hat{{\boldsymbol d}} | g \rangle \; ,
\end{equation} 
with $m$ the electron's mass, 
$q$ its charge, and ${\cal E}_i$ the atomic energy levels
(see, eg., p.~74 of Ref.~\cite{Sak76}).
In the interaction picture with respect to the system
Hamiltonian (\ref{Hs}) 
the momentum operator becomes
\begin{align}
\label{eq:momint}
  \hat{\boldsymbol{p}}_{n}(-\tau) = 
  \hat{\sigma}_{n\textrm{\tiny{+}}} \text{e}^{-\text{i}
  \tau \omega_0 } \boldsymbol{p} +
  \hat{\sigma}_{n\textrm{\tiny{-}}} \text{e}^{\text{i}\tau \omega_0 } 
  \boldsymbol{p}^* ,
\end{align}   
where $\hat{{\boldsymbol p}}_{n} (\tau) = U^{\dagger}_{\text{s}} 
\hat{{\boldsymbol p}}_{n} U_{\text{s}}$
and $U_{\text{s}} = \text{exp}(-\text{i} \tau \widehat{H}_{\text{s}} /\hbar)$.
For convenience, $\hat{{\boldsymbol
    p}}_{n} (0) = \hat{{\boldsymbol p}}_{n}$. 
The interaction Hamiltonian
describes a minimal-coupling between the atom and electric field
\begin{align}
\widehat{H}_{\text{int}} =  -\frac{q}{m}
  \sum_{n=1}^N \hat{\boldsymbol{p}}_n \cdot
  \widehat{\boldsymbol{A}}(\boldsymbol{r}_n) +  \frac{q^2}{2m} \sum_{n=1}^N
  \widehat{\boldsymbol{A}}(\boldsymbol{r}_n)^2,
\end{align}
where $\widehat{\boldsymbol{A}}(\boldsymbol{r}_n)$ is the Hermitian
operator of the vector potential.  
The full
Hamiltonian in the Coulomb gauge, neglecting the free, transverse
electric-field and the free Hamiltonian of atom $n$, 
\begin{align}
\label{eq:hmc}
\widehat{H}_{\text{mc}} = \widehat{H}_{\text{s}} + \widehat{H}_{\text{int}} + 
\widehat{H}_{\text{c}} \; ,
\end{align}
also includes a static
interaction described by~\cite{Pow57,Fri73,Aga74}
\begin{align}
\label{eq:grodip}
\widehat{H}_{\text{c}} = \sum_{n,m = 1}^N  \hbar \Delta^\parallel_{nm}
\hat{\sigma}_{n+}\hat{\sigma}_{m-}, 
\end{align}
for $\Delta^\parallel_{nm}$ of Eq.~(\ref{eq:deltaParUnreg}).
The electrostatic dipole-dipole interaction 
describes a nonretarded
interaction that is not mediated by a transverse 
propagator~\cite{Pow57,Pow59,Aga74}.  The term proportional to
$\widehat{{\boldsymbol A}}^2$ can be omitted because in the dipole
approximation it contributes the same energy to every atomic
state~\cite{Mil94}.  This term includes 
the self-energy of the electric dipole, which also does not contribute to
relative energy shifts~\cite{Coh89}.  So, 
\begin{align}
\label{eq:hintmc}
\widehat{H}_{\text{int}} = -\frac{q}{m} \sum_{n=1}^N \hat{\boldsymbol{p}}_n
\cdot \widehat{\boldsymbol{A}}(\boldsymbol{r}_n) = -\frac{q}{m}
\sum_{i = 1}^3 \sum_{n=1}^N  \hat{p}_{n,i}\widehat{A}_i (\boldsymbol{r}_n),
\end{align}
where the subscript $i$ labels the $i^{\text{th}}$ vector component.
In the Born-Markov approximation, the time evolution of the reduced density
operator for identical 2LAs $n$ and $m$ is given by     
\begin{align}
\label{eq:memc}
 \dot{\rho} =
  &-\frac{\text{i}}{\hbar}[\widehat{H}_{\text{s}} + 
  \widehat{H}_{\text{c}}, \rho]  
\nonumber \\
  &-\frac{q^2}{m^2\hbar^2} \int_0^\infty \text{d}\tau 
  \sum_{i,j = 1}^3\sum_{n,m=1}^N 
 \big( \langle \widehat{A}_i(\tau,\boldsymbol{r}_n)
 \widehat{A}_j(0,\boldsymbol{r}_m) \rangle \nonumber \\
& \times [\hat{p}_{n,i}, \hat{p}_{m,j}(-\tau)
   \rho ] + \langle \widehat{A}_j(0,\boldsymbol{r}_m)
 \widehat{A}_i(\tau,  \boldsymbol{r}_n) \rangle \nonumber \\
&\times [\rho  \hat{p}_{m,j}(-\tau), \hat{p}_{n,i}] 
\big),  
\end{align}  
for $\widehat{A}(\tau) = U^{\dagger}_{\text{R}}
\widehat{A} U_{\text{R}}$ for $U_{\text{R}} = \text{exp}(-\text{i} \tau
\widehat{H}_{\text{R}})$ the reservoir operators in the interaction picture.
The reservoir Hamiltonian $\widehat{H}_R$ is time independent, so $\langle 
\widehat{A}_i(\tau,\boldsymbol{r}_n) \widehat{A}_j(0,\boldsymbol{r}_m) \rangle
= \langle \widehat{A}_j(0, \boldsymbol{r}_m)
\widehat{A}_i(-\tau,\boldsymbol{r}_n) \rangle$.  Making the rotating-wave
approximation and assuming that the reservoir is initially in the vacuum
state, the master equation can be written  
\begin{align}
\dot{\rho} =
  &-\frac{\text{i}}{\hbar}[\widehat{H}_{\text{s}} +  \widehat{H}_{\text{c}},
    \rho] + \sum_{n=1}^N F^{-}_{nn} [ \hat{\sigma}_{nz},\rho]
  \nonumber \\ 
&- \sum_{n,m = 1}^N I_{nm}^+ [\hat{\sigma}_{n+},\hat{\sigma}_{m-}
    \rho ] + (I_{mn}^+)^* [\rho \hat{\sigma}_{m+},\hat{\sigma}_{n-} ],
\end{align}
where $I_{nm}^\pm \equiv F_{nm}^+ \pm  F_{nm}^-$ for
\begin{align}
\label{eq:fpm}
F_{nm}^{\pm} \equiv& \sum_{i,j = 1}^3
  \frac{q^2 p_i p^*_j}{m^2 \hbar^2} 
 \lim_{\epsilon \rightarrow 0} \int_0^\infty \text{d}\tau 
\times \bra{0} \widehat{A}_i^{(+)}(\tau,\boldsymbol{r}_n)\nonumber \\
& \times \widehat{A}_j^{(-)}(0,\boldsymbol{r}_m)  
  \ket{0} \text{e}^{\pm \text{i} \omega_0 \tau - \epsilon \tau} \nonumber \\
=&\sum_{i,j = 1}^3 \frac{\omega_0^2 d_i d_j^*}{\hbar^2} 
  \lim_{\epsilon \rightarrow 0} \int_0^\infty \text{d}\tau
\bra{0} \widehat{A}_i^{(+)}(\tau,\boldsymbol{r}_n) \nonumber \\
& \times \widehat{A}_j^{(-)}(0,\boldsymbol{r}_m)  
  \ket{0} \text{e}^{\pm \text{i} \omega_0 \tau - \epsilon \tau} ,
\end{align}
in which the infinitesimal number $\epsilon > 0$ has been added to ensure the
correct analytical properties in the complex plane.  
In the Coulomb gauge, the
electric field is separated into propagating (transverse) and nonpropagating
(longitudinal) parts.  Here, the superscript `$\perp$' labels transverse in
order to distinguish between the propagating (transverse)
and static (longitudinal) dipole-dipole interaction.  The correlation function
is 
\begin{align}
\label{eq:cor1mc}
\langle \widehat{A}_i^{(+)}(\tau,\boldsymbol{r}_n)&
  \widehat{A}_j^{(-)}(0,\boldsymbol{r}_m) \rangle = 
  \frac{\hbar }{4 \pi^2 \varepsilon_0 c}   \int_0^\infty \text{d}k \; 
  k \, \nonumber \\
&\times \text{e}^{-\text{i} \omega_k \tau }
 ( \alpha(k r_{nm}) + \vec{r}_{nm,i}\vec{r}_{nm,j}
  \beta(k r_{nm})), 
\end{align}
where the frequency of the electric field $\omega_k =c k$ for $k =
|{\boldsymbol  k}|$, and $\alpha$ and $\beta$ given in Eq.~(\ref{eq:alpha})
and (\ref{eq:beta}), respectively.
We use an arrow to denote a unit vector. For instance,
$\vec{r}_{nm}$ corresponds to the unit vector $\vec{r}_{nm}=
\boldsymbol{r}_{nm} / r_{nm}$ in the direction of the vector
$\boldsymbol{r}_{nm} \equiv \boldsymbol{r}_n -\boldsymbol{r}_m$.
In order to calculate 
explicit expressions for $\gamma_{nm}$ and $\Delta_{nm}$, $F_{nm}^\pm$ is
written as   
\begin{align} 
\label{eq:fnmmc}
  F_{nm}^\pm &=   -\text{i}
  \frac{\omega_0^2 |{\boldsymbol d}|^2}{4 \pi^2 \varepsilon_0 \hbar c} 
  \lim_{\epsilon \rightarrow 0} 
  \int_0^\infty \text{d}k \; 
  k \,  \frac{ \alpha(k r_{nm})+ \eta_{nm} \beta(k r_{nm}) }{\omega_k \mp
  \omega_0 - 
  \text{i}\epsilon} ,
\nonumber \\ &=
\frac{ 3\gamma}{4\pi \text{i} k_0}\lim_{\epsilon \rightarrow 0} 
  \int_0^\infty \text{d}k\; k \,
   \frac{\alpha(k r_{nm}) +\eta_{nm} \beta(k r_{nm})}{k \mp k_0 -
  \text{i}\epsilon} \; , 
\end{align} 
So, the quantity $I_{nm}^+$
can be written 
\begin{align}
\label{eq:fnmmc1}
 I_{nm}^+= 
 \frac{ 3\gamma}{4\pi \text{i} k_0} \lim_{\epsilon \to 0}
  \int_{-\infty}^\infty \text{d}k\; k \,
  \frac{\alpha(k r_{nm})+\eta_{nm} \beta(k r_{nm})}{k - k_0 - \text{i}\epsilon}
  \; ,
\end{align} 
which can be integrated using the
residue theorem.   It has a pole at $k_0 + \text{i}\epsilon$, and so
is integrated taking care to separate terms proportional to $\exp (\text{i} k
r_{nm})$, for which the contour has to  
be closed in the upper half plane, from those terms proportional to 
$\exp (-\text{i} k r_{nm})$.  While the total expression has no pole at $k=0$,
each of these terms diverges  at this point and the corresponding residue
has to be taken into account. 
Eq.~\eqref{eq:fnmmc1} then leads to 
$\gamma_{nm} \equiv \Re (I_{nm}^+)$ given in Eq.~(\ref{eq:mcurg}) and
$\Delta_{nm}^\perp \equiv \Im ( I_{nm}^+ )$ given in Eq.~(\ref{eq:rtddnr}). 
The collective emission and level shift coefficients are identified as 
$\gamma_{nm}$ and $\Delta_{nm}^\perp + \Delta_{nm}^\parallel$, 
respectively. The last term in 
Eq.~\eqref{eq:rtddnr}, which arises from the pole at $k = 0$, 
cancels with the static part of the full Hamiltonian quoted in
Eq.\eqref{eq:grodip}~\cite{Pow57,Pow59,Aga74,Fri73}.  So, the master equation
is written
\begin{align}
\label{eq:meun}
 \dot{\rho} =& -\frac{\text{i}}{\hbar}[\widehat{H}_{\text{s}}, \rho]  
-\text{i} \sum_{n,m = 1}^N
  (\Delta^\perp_{nm} + \Delta^\parallel_{nm}
  )  [\hat{\sigma}_{n+} \hat{\sigma}_{m-}, \rho] \nonumber \\ 
&+ \sum_{n,m = 1}^N
  {\gamma}_{nm} 
  (2 \hat{\sigma}_{m-} \rho \hat{\sigma}_{n+} - \hat{\sigma}_{n+}
  \hat{\sigma}_{m-} \rho - \rho 
  \hat{\sigma}_{n+} \hat{\sigma}_{m-} ).   
\end{align}

The single-atom level shift obtained from the master equation for a
single-atom 
\begin{align}
\label{eq:sing}
\dot{\rho} =&
-\frac{\text{i}}{\hbar}[\widehat{H}_{\text{s}} +  \widehat{H}_{\text{c}},
    \rho] - \frac{\text{i}}{2} \Im \{ I_{nn}^- \} [\hat{\sigma}_z,\rho]
\nonumber \\
&-
\Re\{ I_{nn}^+ \} (\hat{\sigma}_+  \hat{\sigma}_-  \rho + \rho \hat{\sigma}_+
\hat{\sigma}_- - 2  \hat{\sigma}_-  \rho \hat{\sigma}_+ ),
\end{align}
is equal to the imaginary part of $I_{nn}^-$
\begin{align}
\label{eq:innmc1}
\Delta_\text{Lamb}^{\text{(2-lev)}} 
 & \equiv \Im\{I_{nn}^-\}
\nonumber \\
&= \frac{ \gamma}{2 \pi k_0} \left( \mathcal{P} 
\int_0^\infty \text{d}k
\frac{ k}{k
  - k_0}  -\mathcal{P}  \int_0^\infty \text{d}k \frac{ k}{k +  k_0}  \right),
\nonumber   \\ 
&= \frac{\gamma}{2 \pi } \mathcal{P} \int_0^\infty  \text{d}k \left(
\frac{1}{k - 
  k_0} + \frac{1}{k + k_0} \right),
\end{align}
where the limit $\xi_{nm} \to 0$ has been taken in $F^\pm_{nm}$, and where
$\mathcal{P}$ denotes principal value. This normally is absorbed into
$\widehat{H}_s$ by appropriate 
redefinition of the eigenfrequency $\omega_0$.  It is connected with the
single-atom Lamb shift~\cite{Bet47}, but 
because of the two-level approximation it only includes the 
energy shift generated by the coupling between the two basis states.

\subsection*{Regularization}
\label{sec:remc}
The divergence of both Eq.~\eqref{eq:fnmmc1} and Eq.~\eqref{eq:innmc1}
can be removed using a regularization factor
$\Lambda_\perp^2/(\Lambda_\perp^2 + k^2)$.  For the imaginary part of the
single-atom quantity, this gives
\begin{align}
\label{eq:innmc2}
\Delta_\text{Lamb}^{\text{(2-lev)}}  &=  \frac{\gamma}{2 \pi } \mathcal{P}
  \int_0^\infty 
  \text{d}k 
  \left( \frac{1}{k - 
  k_0} + \frac{1}{k + k_0} \right) \frac{  \Lambda_\perp^2}{k^2 +
  \Lambda_\perp^2}.
\end{align}
This integral becomes
\begin{align}
 \label{eq:innmc3}
 \Delta_\text{Lamb}^{\text{(2-lev)}} =  \frac{ \tilde{\gamma} }{\pi} 
         \ln \left (\frac{\Lambda_\perp}{k_0} \right). 
\end{align} 
The multi-atom quantity $I_{nm}^+$ is regularized in a similar way
\begin{align}
\label{eq:inmmc}
 \tilde{I}_{nm}^+  =  \frac{ 3\gamma}{4\pi \text{i} k_0}\lim_{\epsilon \to 0}
  \int_{-\infty}^\infty &\text{d}k\; k\,
  \frac{\alpha(k r_{nm})+\eta_{nm} \beta(k r_{nm})}{k - k_0 - \text{i}\epsilon}
 \nonumber \\
&\times \frac{\Lambda_\perp^2}{\Lambda_\perp^2 + k^2}. 
\end{align}
The small $r_{nm}$ behaviour is now moderated. By implication, the details of
the interaction region are not of specific interest~\cite{Vrie}.  
We remark that
this regularization procedure is not unique.  Eq.~\eqref{eq:inmmc} has poles
at $k_0 + \text{i}\epsilon$ and $\pm \text{i} \Lambda_\perp$, and so is
integrated using the residue theorem. Using
$\tilde{\gamma}$ of Eq.~(\ref{eq:gammaTilde})
and calculating the integral in Eq.~\eqref{eq:inmmc} results in
Eqs.~(\ref{eq:mcrg}) and (\ref{eq:mcrd}), where we defined
$ \tilde{\gamma}_{nm} \equiv  \Re(\tilde{I}_{nm}^+)$ and 
$ \tilde{\Delta}^\perp_{nm} \equiv \Im(\tilde{I}_{nm}^+)$.

Taking $\Lambda_\perp \to \infty$ recovers Eq.~\eqref{eq:mcurg}
and~\eqref{eq:rtddnr}.  The dipole-dipole interaction 
$\tilde{\Delta}_{nm}^\perp$ is now regularized, with the limiting values given
by 
\begin{align}
 \lim_{r_{nm} \to 0} \tilde{\Delta}^\perp_{nm} =
  -\frac{\tilde{\gamma} \Lambda_\perp}{2 k_0}, 
\end{align}
and 
\begin{align}
\lim_{r_{nm} \to 0} \tilde{\gamma}_{nm} =
  \frac{\tilde{\gamma}}{2}.  
\end{align}       
Note that if $\Lambda_\perp \gg k_0$, $\tilde{\gamma} \simeq \gamma$, and also
that $\lim_{r_{nm} \to 0} \gamma_{nm} = \gamma/2$, 
where $\gamma_{nm}$ is
defined in Eq.~\eqref{eq:mcurg}.

For $\Lambda_\perp \to \infty$, the final term in Eq.~\eqref{eq:mcrd} cancels
the static dipole-dipole interaction to give Eq.~\eqref{eq:meun}.
However, subtracting this term as before changes the properties of
$\tilde{\Delta}_{nm}^\perp$ such that it once again diverges as $r_{nm} \to
0$.  So, in order to maintain the analytic properties of $\tilde{I}_{nm}^+$ in
the limit $r_{nm} \to 0$, it is necessary to retain the final term in
$\tilde{\Delta}_{nm}^\perp$.  The problem is then that the static,
divergent, dipole-dipole interaction $\widehat{H}_c$ of Eq.~(\ref{eq:grodip})
remains in the original Hamiltonian.
In order to account for the divergence, we regularize 
$\widehat{H}_c$ as follows.  
First, notice that~\cite{Coh89} 
\begin{align}
\Delta^\parallel_{nm} = \frac{|\boldsymbol{d}|^2}{ \hbar \varepsilon_0} \left(
\vec{d}_n \cdot G(k,\boldsymbol{r}) \cdot \vec{d}_m \right),
\end{align}
where the Green's function $G(k,\boldsymbol{r})$ is  
\begin{align}
\label{eq:stat1a}
G_{ij}(k,\boldsymbol{r}) = \int \frac{\text{d}^3 k}{(2 \pi)^3}
  \vec{k}_i \vec{k}_j \text{e}^{\text{i} \boldsymbol{k} \cdot 
    (\boldsymbol{r}_n  - \boldsymbol{r}_m)} .
\end{align}  
This can be regularized in a similar way to before: 
\begin{align}
\label{eq:stat1}
\tilde{G}_{ij}(k,\boldsymbol{r}) =  \int 
\frac{\text{d}^3 k}{(2 \pi)^3}  \vec{k}_i \vec{k}_j
\text{e}^{\text{i} 
    \boldsymbol{k} \cdot 
    (\boldsymbol{r}_n  - \boldsymbol{r}_m)} 
\frac{\Lambda^4_\parallel}{k^4 + \Lambda^4_\parallel}.  
\end{align}
The regularization parameter is raised to the fourth power in order to account
for the $r^3$ divergence of the static interaction.  Eq.~\eqref{eq:stat1} can
be written 
\begin{align}
\tilde{\Delta}^\parallel_{nm} =& \frac{|\boldsymbol{d}|^2}{2 \hbar
 \pi^2 \varepsilon_0} \int_0^\infty k^2 \text{d} k
\Bigg\{ \frac{\sin \xi_{nm}}{ \xi_{nm}^3} - \frac{\cos \xi_{nm} }{\xi_{nm}^2}
\nonumber \\
&+ \eta_{nm} \left[ \frac{3 \cos \xi_{nm}}{\xi_{nm}^2} - \frac{3 \sin
    \xi_{nm}}{\xi_{nm}^3} +  \frac{\sin \xi_{nm}}{\xi_{nm}} 
\right] \Bigg\} \nonumber \\
&\times \frac{\Lambda^4_\parallel}{k^4 + \Lambda^4_\parallel}.
\end{align}
Evaluating this expression gives Eq.~(\ref{eq:delp}).
Again, taking $\Lambda_\parallel \to \infty$ recovers $\Delta^\parallel_{nm}$.
The static interaction is now regularized with limiting value given by
\begin{align}
 \lim_{r_{nm} \to 0}
\tilde{\Delta}^\parallel_{nm} = 
 \frac{ \gamma \Lambda_\parallel^3}{4 \sqrt{2} k_0^3}.
\end{align}
Hence, the fully regularized master equation is given by Eq.~(\ref{eq:mefmc}),
where the parameters are given by $\tilde{\gamma}_{nm}$ of Eq.~(\ref{eq:mcrg}),
$\tilde{\Delta}^\perp_{nm}$ of Eq.~(\ref{eq:mcrd}), and
$\tilde{\Delta}^\parallel_{nm}$ of Eq.~(\ref{eq:delp}).  In the regime where
regularization matters, there is no longer a complete cancellation of
$\widehat{H}_c$.  As the separation of the 2LAs increases,
Eq.~\eqref{eq:mefmc} 
approaches Eq.~\eqref{eq:meun}.  The regularization has
smeared the zero-spatial extent of the point-dipoles.
It remains to show the equivalence with the electric-dipole Hamiltonian and to
propose a value for $\Lambda_\parallel$ and $\Lambda_\perp$.  Here,
we note a few important remarks.

First, as is normal in the Coulomb gauge, the electric field is split into a
longitudinal and a transverse field.  The retarded nature of the
electromagnetic field results from an exact compensation between two
instantaneous parts derived from the Coulomb field and the transverse field
respectively (see Eq.~\eqref{eq:rtddnr}).  With the cut-offs applied, this
cancellation no longer occurs in the near-Dicke limit. The dipole-dipole
interaction at small separations includes, but does not consist entirely of, a
contact term.  This is in contrast to the result, stated in
Eq.~\eqref{eq:grodip}, derived in Ref.~\cite{Gro}.
We conjecture that the origin of non-retarded terms lies in the
choice of different cut-off parameters for transverse and 
longitudinal fields, which explicitly breaks the
covariance of Maxwell's theory and hence can violate causality.
The ultimate reason for this different treatment is that we work
in Coulomb gauge, which is the standard choice to describe atom-light
interactions but which is not covariant.
It would probably be possible to remove the non-retarded terms
by using a covariant description of the atom-light coupling,
but this effort would not be justified because in the near-Dicke limit
retardation effects can safely be ignored.

Second, the assumptions behind our model imply that 
relativistic effects, such as vacuum polarization or relativistic recoil,
are not included. This approximation is justified as long as the
resonance energy $\hbar \omega_0$ of the two-level systems is much
smaller than the rest energy $m c^2$ of the electron. Relativistic
effects would lead to modifications of our treatment on a length
scale of the order of the electron's Compton wavelength $2\pi\hbar/(m c)$.
\section{Derivation of the regularized master equation
in electric-dipole coupling}
\label{sec:reged} 
The Lindblad master equation derived previously describes the
evolution of a collection of 
minimally coupled 2LAs.  We would expect
the result to be the same using the multipolar Hamiltonian for two reasons.
First, Eqs.~\eqref{eq:mcurg} and~\eqref{eq:rtddnr} are equivalent to the
more commonly used electric-dipole master 
equation~\cite{Bela69,Lehm70i,Arg70,Car00,Clem03}.  Second, the 
 full minimal-coupling Hamiltonian is unitarily equivalent to the full
 multipolar Hamiltonian. We follow the same method as
Sec.~\ref{sec:regmin}, but take care to distinguish the differences.
Transforming to the interaction picture gives
\begin{align}
\label{eq:dipint}
  \hat{\boldsymbol{d}}_{n}(-\tau) = 
  \widehat{\sigma}_{n\textrm{\tiny{+}}} \text{e}^{-\text{i}
  \tau \omega_0 } \boldsymbol{ d} +
  \widehat{\sigma}_{n\textrm{\tiny{-}}} \text{e}^{\text{i}\tau \omega_0 } 
  \boldsymbol{ d}^* ,
\end{align}   
where $\hat{\boldsymbol{ d}}_{n} (\tau) = U^{\dagger}_{\text{s}} 
\hat{\boldsymbol{ d}}_{n} U_{\text{s}}$.  For convenience, $\hat{{\boldsymbol
    d}}_{n} (0) = \hat{{\boldsymbol d}}_{n}$. The interaction Hamiltonian
describes an electric-dipole coupling between the atom and electric field
\begin{align}
\widehat{H}_{\text{int}} = -\sum_{n=1}^N \hat{\boldsymbol{d}}_n \cdot
\widehat{\boldsymbol{E}}(\boldsymbol{r}_n) = - \sum_{i =1}^3 \sum_{n=1}^N
\hat{d}_{n,i} \widehat{E}_i(\boldsymbol{r}_n).
\end{align}
Using the multipolar Hamiltonian, an extra term
\begin{align}
\label{eq:dipself}
\widehat{H}_{\text{self}} = \frac{1}{2 \varepsilon_0} \sum_{n\ne m=1}^N \int
\text{d}^3 \boldsymbol{r} 
   \widehat{\boldsymbol{P}}_n(\boldsymbol{r}) \cdot
   \widehat{\boldsymbol{P}}_m(\boldsymbol{r}) ,  
\end{align}
where, in the electric-dipole approximation,
\begin{align}
 \widehat{\boldsymbol{P}}_n(\boldsymbol{r}) = \hat{\boldsymbol{d}}_{n}
 \delta(\boldsymbol{r} - \boldsymbol{r}_n) 
\end{align}
describing atomic self-energies and contact interactions is present.  These
terms result from applying the Power-Zienau-Woolley transformation
\begin{align}
\mathcal{U} = \text{exp}\left[-\frac{\text{i}}{\hbar} \sum_{n=1}^N \int
  \text{d}^3 \boldsymbol{r} \widehat{\boldsymbol{P}}_n(\boldsymbol{r}) 
  \cdot \widehat{\boldsymbol{A}}(\boldsymbol{r}_n)\right]
\end{align}
to Eq.~\eqref{eq:hmc}.  We apply the transformation as $\mathcal{U}
\widehat{H}_{\text{mc}} \mathcal{U}^\dagger$, so $\widehat{\boldsymbol{E}}$
represents the electric-field.  See Ref.~\cite{Coh89} for an introduction to
the Power-Zienau-Woolley transformation, and Ref.~\cite{Dav75} for a deeper
analysis that refutes some of the results in Ref.~\cite{Pow57} and highlights
the importance of the order with which one applies the transformation.  We
assume that 
any terms that refer to self-energies of a single atom are renormalized into
$\widehat{H}_{\text{s}}$. Thus, the full electric-dipole Hamiltonian is  
\begin{align}
\label{eq:hed}
\widehat{H}_{\text{ed}} = \widehat{H}_{\text{s}} + \widehat{H}_{\text{int}} + 
\widehat{H}_{\text{self}}.
\end{align}  
In the Born-Markov and electric-dipole approximation, the time evolution of
the reduced density operator for atoms $n$ and $m$ is given by  
\begin{align}
\label{eq:medc}
\dot{\rho} =& 
-\frac{\text{i}}{\hbar}[\widehat{H}_{\text{s}} +
\widehat{H}_{\text{self}},\rho]  \nonumber \\
& -\frac{1}{\hbar^2} \int_0^\infty \text{d}\tau 
\sum_{i,j=1}^3  \sum_{n,m=1}^N 
 \big( \langle \widehat{E}_i(\tau,\boldsymbol{r}_n)
 \widehat{E}_j(0,\boldsymbol{r}_m) \rangle \nonumber \\
& \times [\hat{d}_{n,i}, \hat{d}_{m,j}(-\tau)
   \rho ] \nonumber \\
&+ \langle \widehat{E}_j(0,\boldsymbol{r}_m)
 \widehat{E}_i(\tau,  \boldsymbol{r}_n) \rangle [\rho  
\hat{d}_{m,j}(-\tau), \hat{d}_{n,i}] 
\big),  
\end{align}
for $\widehat{E}(\tau) = U^{\dagger}_{\text{R}}
\widehat{E} U_{\text{R}}$ 
for $U_{\text{R}} = \text{exp}(-\text{i} \tau \widehat{H}_{\text{R}})$ the
reservoir operators in the interaction picture. The reservoir Hamiltonian
$H_R$ is time independent, so $\langle
\widehat{E}_i(\tau,\boldsymbol{r}_n)
\widehat{E}_j(0,\boldsymbol{r}_m) \rangle = \langle 
\widehat{E}_j(0, \boldsymbol{r}_m) 
\widehat{E}_i(-\tau,\boldsymbol{r}_n) \rangle$. 
Following the notation of Sec.~\ref{sec:regmin}, $F_{nm}^\pm$ is defined as  
\begin{align}
\label{eq:fped}
F_{nm}^{\pm} \equiv& \sum_{i,j =1}^3
  \frac{d_i d_j^*}{\hbar^2} 
 \lim_{\epsilon \rightarrow 0} \int_0^\infty \text{d}\tau 
\bra{0} \widehat{E}_i^{(+)}(\tau,\boldsymbol{r}_n) \nonumber \\
&\times \widehat{E}_j^{(-)}(0,\boldsymbol{r}_m)  
  \ket{0} \text{e}^{\pm \text{i} \omega_0 \tau - \epsilon \tau}
\end{align}
and the correlation function is 
\begin{align}
\label{eq:cor1ed}
\langle \widehat{E}_i^{(+)}(\tau,\boldsymbol{r}_n)&
  \widehat{E}_j^{(-)}(0,\boldsymbol{r}_m) \rangle = 
  \frac{\hbar c}{4 \pi^2 \varepsilon_0 }   \int_0^\infty \text{d}k \; 
  k^3 \, \nonumber \\
&\times \text{e}^{-\text{i} \omega_k \tau }
 ( \alpha(\xi_{nm}) + \vec{r}_{nm,i}\vec{r}_{nm,j}
  \beta(\xi_{nm})).
\end{align} 
This is not equal to, but has the same fundamental 
properties as Eq.~\eqref{eq:cor1mc} which means the master
equation can be written in the form
\begin{align}
\label{eq:meff}
\dot{\rho } =&
  -\frac{\text{i}}{\hbar}[\widehat{H}_{\text{s}} +  \widehat{H}_{\text{self}},
    \rho] 
+  \sum_{n=1}^N F^{-}_{nn} [ \hat{\sigma}_{nz},\rho]
\nonumber \\
& -  \sum_{n,m = 1}^N I_{nm}^+ [\hat{\sigma}_{n+},\hat{\sigma}_{m-}
    \rho ] + (I_{mn}^+)^* [\rho \hat{\sigma}_{m+},\hat{\sigma}_{n-} ],
\end{align}
where $I_{nm}^\pm \equiv F_{nm}^+ \pm  F_{nm}^-$.  So, as in 
Sec.~\ref{sec:regmin} the collective emission
coefficient and the level shift operator are identified as 
$\gamma_{nm} \equiv \Re ( I_{nm}^+ )$ and $\Delta_{nm} \equiv \Im
(I_{nm}^+ )$ respectively.  The non-regularized quantity $F_{nm}^\pm$
is written as  
\begin{align} 
\label{eq:fned}
  F_{nm}^\pm = \frac{|\boldsymbol{ d}|^2}{4\pi^2 \text{i} \varepsilon_0
  \hbar} \lim_{\epsilon \to 0} \int_0^\infty \text{d}k \; k^3 
  \frac{\alpha(\xi_{nm}) + \eta_{nm}  \beta(\xi_{nm}) }{k \mp k_0 -
  \text{i}\epsilon}.   
\end{align} 
The crucial difference between Eq.~\eqref{eq:fnmmc} and
Eq.~\eqref{eq:fned} is the order of the divergence.  Here, it is 
proportional to $k^3$ but in minimal-coupling it is proportional to $k$.
If this divergence is not accounted for and the regularization proceeds
directly from here, the two regularized answers using the minimal-coupling and
the electric-dipole Hamiltonians will be different.  
\subsection*{Equivalence with minimal-coupling}
\label{sec:equivmc} 
In order to account for the difference in the divergence, the self-energy
given by Eq.~\eqref{eq:dipself} is examined more closely.  Calculating this
integral sheds light on the origin of the higher divergence of
Eq.~\eqref{eq:fned}: in the electric-dipole picture the self-energies of the
atomic dipoles have been implicitly included.  These same energies form part
of the $\widehat{{\boldsymbol A}}^2$ term in minimal-coupling.  So,
making the rotating wave approximation and using the commutation relation
$[\hat{\sigma}_{n-},\hat{\sigma}_{m+}] = 0 \; \forall \; n \ne m$, the
self-energy contribution is   
\begin{align}
\widehat{H}_{\text{self}} = \sum_{i,j=1}^3 
\frac{d_i\, \delta_{ij}\,  d_j^*}{ \varepsilon_0}
\sum_{n \ne m=1}^N \hat{\sigma}_{n+}\hat{\sigma}_{m-} 
  \delta(\boldsymbol{r}_n - \boldsymbol{r}_m),
\end{align} 
which can be decomposed into longitudinal and transverse parts
\begin{align}
\widehat{H}_{\text{self}} =& \sum_{i,j=1}^3 \frac{d_i d_j^* }{ \varepsilon_0}
\sum_{n \ne m=1}^N \hat{\sigma}_{n+}\hat{\sigma}_{m-} \nonumber \\
&\times  \bigg\{ \delta^\perp_{ij}(\boldsymbol{r}_n - \boldsymbol{r}_m) 
+
  \delta^\parallel_{ij}(\boldsymbol{r}_n - \boldsymbol{r}_m) \bigg\}.
\end{align} 
The transverse delta function, which is proportional to the commutator of the
vector potential and the transverse electric field, is defined by
\begin{align}
    \delta^\perp_{ij}(\boldsymbol{r}_n - \boldsymbol{r}_m) & = 
    \int \frac{\text{d}^3
    k}{(2 \pi)^3}  \left(\delta_{ij} -\frac{{k}_i
    {k}_j}{k^2}\right) 
    \text{e}^{\text{i} \boldsymbol{k} \cdot 
    (\boldsymbol{r}_n  - \boldsymbol{r}_m)}
\nonumber 
\\ & = 
\frac{1}{2 \pi^2 } \int_0^\infty \text{d}k \;
k^2 \, \{ \alpha(\xi_{nm}) \, \delta_{ij}
\nonumber \\ & \hspace{2cm}
  + \vec{r}_{nm,i}\, \vec{r}_{nm,j}\, 
  \beta(\xi_{nm}) \}
\end{align}
for $\alpha(\xi_{nm})$ and $\beta(\xi_{nm})$ defined in Eqs.~\eqref{eq:alpha}
and~\eqref{eq:beta}, respectively. 
The longitudinal delta function is defined as 
\begin{align}
\label{deltaPar}
 \delta^\parallel_{ij}(\boldsymbol{r}_n - \boldsymbol{r}_m) 
 = \int \frac{\text{d}^3
    k}{(2 \pi)^3}  \frac{  k_i k_j }{k^2}
    \text{e}^{\text{i} \boldsymbol{k} \cdot 
    (\boldsymbol{r}_n  - \boldsymbol{r}_m)} \; ,
\end{align}
which is the same expression as in  Eq.~\eqref{eq:stat1a}.
The transverse part of
$\widehat{H}_\text{self}$ becomes 
\begin{align}
\widehat{H}^\perp_{\text{self}} =& \frac{|\boldsymbol{d}|^2}{ 2 \pi^2
  \varepsilon_0} \sum_{n \ne m=1}^N \hat{\sigma}_{n+}\hat{\sigma}_{m-} 
 \int_0^\infty \text{d}k \;
k^2 \{ \alpha(\xi_{nm}) \nonumber \\
&+ \eta_{nm}  \beta(\xi_{nm}) \}
 \end{align}
for $\eta_{nm} \equiv ( \vec{d} \cdot \vec{r}_{nm})^2$.  
Writing
$I_{nm}^+$ in electric-dipole coupling as
\begin{align}
I_{nm}^+ &= \frac{|\boldsymbol{ d}|^2}{4\pi^2 \text{i} \varepsilon_0 
  \hbar} \int_{-\infty}^\infty \text{d}k \; k^3 \frac{ \alpha(\xi_{nm}) +
  \eta_{nm} 
  \beta(\xi_{nm}) }{k - k_0 - \text{i}\epsilon} \nonumber \\
&= \frac{|\boldsymbol{ d}|^2}{4\pi^2 \text{i} \varepsilon_0 
  \hbar}\bigg\{ \int_0^\infty \text{d}k \; 2 k^2 \{ \alpha(\xi_{nm}) +
  \eta_{nm} 
  \beta(\xi_{nm}) \} \nonumber \\
&\phantom{xxxxx} + k_0 \int_{-\infty}^\infty \text{d}k \; k^2 
\frac{ \alpha(\xi_{nm}) + \eta_{nm}
  \beta(\xi_{nm}) }{k - k_0 - \text{i}\epsilon}  \bigg\},
\end{align}   
it can be seen that, when written as part of the master equation
\eqref{eq:meff}, the first term, multiplied by the relevant operators, cancels
$\widehat{H}^\perp_{\text{self}}$.  After some algebra, the remainder of 
$I_{nm}^+$ can be written
\begin{align} 
\label{eq:inmed}
I_{nm}^+= 
 \frac{ 3\gamma}{4\pi \text{i} k_0} \lim_{\epsilon \to 0}
  \int_{-\infty}^\infty \text{d}k\; k \,
  \frac{\alpha(k r_{nm})+\eta_{nm} \beta(k r_{nm})}{k - k_0 - \text{i}\epsilon}
  \; .
\end{align} 
This is the same as $I_{nm}^+$ stated in Eq.~\eqref{eq:fnmmc1} in
Sec.~\ref{sec:regmin}.  So, by accounting for the transverse polarization in
the electric-dipole Hamiltonian the divergence of the correlation function has
been reduced from $k^3$ to $k$, and the electric-dipole description of the
dipole-dipole interaction has been made identical to that derived in
minimal-coupling.  The transverse polarization squared can be thought of
as the contact interaction between touching, but distinct, 2LAs. 

In the electric-dipole picture, the pole at $k=0$ in the integral of
Eq.~\eqref{eq:inmed} is accounted for (at large separations) by the
longitudinal self-energy contribution.  For small separations,
$\widehat{H}^\parallel_{\text{self}}$ is regularized as follows. Consider
\begin{align}
\widehat{H}^\parallel_{\text{self}} = \sum_{i,j=1}^3 \frac{d_i
  d_j^*}{ \varepsilon_0} \sum_{n \ne m=1}^N \hat{\sigma}_{n+}
  \, \hat{\sigma}_{m-}  \,
    \delta_{ij}^\parallel 
    (\boldsymbol{r}_n - \boldsymbol{r}_m). 
\end{align}  
We can regularize the parallel part of the $\delta$-distribution
of Eq.~(\ref{deltaPar}) as in Eq.~(\ref{eq:stat1a}) to find
\begin{align}
\widehat{H}^\parallel_{\text{self}} = \sum_{n \ne m=1}^N
\tilde{\Delta}^\parallel_{nm} \hat{\sigma}_{n+}\hat{\sigma}_{m-} , 
\end{align}
for $\tilde{\Delta}^\parallel_{nm}$ stated in Eq.~\eqref{eq:delp}.  The master
equation derived using the electric-dipole Hamiltonian is then 
identical to Eq.~\eqref{eq:mefmc}, which has been derived 
using minimal coupling. 
\bibliography{references2-kpm}    

\begin{thebibliography}{31}
\expandafter\ifx\csname natexlab\endcsname\relax\def\natexlab#1{#1}\fi
\expandafter\ifx\csname bibnamefont\endcsname\relax
  \def\bibnamefont#1{#1}\fi
\expandafter\ifx\csname bibfnamefont\endcsname\relax
  \def\bibfnamefont#1{#1}\fi
\expandafter\ifx\csname citenamefont\endcsname\relax
  \def\citenamefont#1{#1}\fi
\expandafter\ifx\csname url\endcsname\relax
  \def\url#1{\texttt{#1}}\fi
\expandafter\ifx\csname urlprefix\endcsname\relax\def\urlprefix{URL }\fi
\providecommand{\bibinfo}[2]{#2}
\providecommand{\eprint}[2][]{\url{#2}}

\bibitem[{\citenamefont{Dicke}(1954)}]{Dicke}
\bibinfo{author}{\bibfnamefont{R.~H.} \bibnamefont{Dicke}},
  \bibinfo{journal}{Phys. Rev.} \textbf{\bibinfo{volume}{93}},
  \bibinfo{pages}{99} (\bibinfo{year}{1954}).

\bibitem[{\citenamefont{Marzlin et~al.}(2007)\citenamefont{Marzlin, Karasik,
  Sanders, and Whaley}}]{kpm07}
\bibinfo{author}{\bibfnamefont{K.-P.} \bibnamefont{Marzlin}},
  \bibinfo{author}{\bibfnamefont{R.}~\bibnamefont{Karasik}},
  \bibinfo{author}{\bibfnamefont{B.~C.} \bibnamefont{Sanders}},
  \bibnamefont{and} \bibinfo{author}{\bibfnamefont{B.~K.}
  \bibnamefont{Whaley}}, \bibinfo{journal}{Can. J. Phys.}
  \textbf{\bibinfo{volume}{85}}, \bibinfo{pages}{641} (\bibinfo{year}{2007}).

\bibitem[{\citenamefont{Kempe et~al.}(2001)\citenamefont{Kempe, Bacon, Lidar,
  and Whaley}}]{Kem01}
\bibinfo{author}{\bibfnamefont{J.}~\bibnamefont{Kempe}},
  \bibinfo{author}{\bibfnamefont{D.}~\bibnamefont{Bacon}},
  \bibinfo{author}{\bibfnamefont{D.~A.} \bibnamefont{Lidar}}, \bibnamefont{and}
  \bibinfo{author}{\bibfnamefont{K.~B.} \bibnamefont{Whaley}},
  \bibinfo{journal}{Phys. Rev. A} \textbf{\bibinfo{volume}{63}},
  \bibinfo{pages}{042307} (\bibinfo{year}{2001}).

\bibitem[{\citenamefont{Lidar et~al.}(1998)\citenamefont{Lidar, Chuang, and
  Whaley}}]{Lidar98}
\bibinfo{author}{\bibfnamefont{D.~A.} \bibnamefont{Lidar}},
  \bibinfo{author}{\bibfnamefont{I.~L.} \bibnamefont{Chuang}},
  \bibnamefont{and} \bibinfo{author}{\bibfnamefont{K.~B.}
  \bibnamefont{Whaley}}, \bibinfo{journal}{Phys. Rev. Lett}
  \textbf{\bibinfo{volume}{81}}, \bibinfo{pages}{2584} (\bibinfo{year}{1998}).

\bibitem[{\citenamefont{Zanardi and Rasetti}(1997)}]{Zan97a}
\bibinfo{author}{\bibfnamefont{P.}~\bibnamefont{Zanardi}} \bibnamefont{and}
  \bibinfo{author}{\bibfnamefont{M.}~\bibnamefont{Rasetti}},
  \bibinfo{journal}{Phys. Rev. Lett} \textbf{\bibinfo{volume}{79}},
  \bibinfo{pages}{3306} (\bibinfo{year}{1997}).

\bibitem[{\citenamefont{Karasik et~al.}(2007)\citenamefont{Karasik, Marzlin,
  Sanders, and Whaley}}]{kar07}
\bibinfo{author}{\bibfnamefont{R.}~\bibnamefont{Karasik}},
  \bibinfo{author}{\bibfnamefont{K.-P.} \bibnamefont{Marzlin}},
  \bibinfo{author}{\bibfnamefont{B.~C.} \bibnamefont{Sanders}},
  \bibnamefont{and} \bibinfo{author}{\bibfnamefont{B.~K.}
  \bibnamefont{Whaley}}, \bibinfo{journal}{Phys. Rev. A}
  \textbf{\bibinfo{volume}{76}}, \bibinfo{pages}{012331}
  (\bibinfo{year}{2007}).

\bibitem[{\citenamefont{Brooke}(2007)}]{Brooke07}
\bibinfo{author}{\bibfnamefont{P.~G.} \bibnamefont{Brooke}},
  \bibinfo{journal}{Phys. Rev. A} \textbf{\bibinfo{volume}{75}},
  \bibinfo{pages}{022320} (\bibinfo{year}{2007}).

\bibitem[{\citenamefont{Beige et~al.}(2000)\citenamefont{Beige, Huelga, Knight,
  Plenio, and Thompson}}]{Bei99}
\bibinfo{author}{\bibfnamefont{A.}~\bibnamefont{Beige}},
  \bibinfo{author}{\bibfnamefont{S.~F.} \bibnamefont{Huelga}},
  \bibinfo{author}{\bibfnamefont{P.~L.} \bibnamefont{Knight}},
  \bibinfo{author}{\bibfnamefont{M.~B.} \bibnamefont{Plenio}},
  \bibnamefont{and} \bibinfo{author}{\bibfnamefont{R.~C.}
  \bibnamefont{Thompson}}, \bibinfo{journal}{J. Mod. Opt.}
  \textbf{\bibinfo{volume}{47}}, \bibinfo{pages}{401} (\bibinfo{year}{2000}).

\bibitem[{\citenamefont{Brennen et~al.}(2000)\citenamefont{Brennen, Deutsch,
  and Jessen}}]{Br00}
\bibinfo{author}{\bibfnamefont{G.~K.} \bibnamefont{Brennen}},
  \bibinfo{author}{\bibfnamefont{I.~H.} \bibnamefont{Deutsch}},
  \bibnamefont{and} \bibinfo{author}{\bibfnamefont{P.~S.}
  \bibnamefont{Jessen}}, \bibinfo{journal}{Phys. Rev. A}
  \textbf{\bibinfo{volume}{61}}, \bibinfo{pages}{062309}
  (\bibinfo{year}{2000}).

\bibitem[{\citenamefont{Petrosyan and Kurizki}(2002)}]{Pet02}
\bibinfo{author}{\bibfnamefont{D.}~\bibnamefont{Petrosyan}} \bibnamefont{and}
  \bibinfo{author}{\bibfnamefont{G.}~\bibnamefont{Kurizki}},
  \bibinfo{journal}{Phys.Rev.Lett} \textbf{\bibinfo{volume}{89}},
  \bibinfo{pages}{207902} (\bibinfo{year}{2002}).

\bibitem[{\citenamefont{Kiffner et~al.}(2007)\citenamefont{Kiffner, Evers, and
  Keitel}}]{Kif07}
\bibinfo{author}{\bibfnamefont{M.}~\bibnamefont{Kiffner}},
  \bibinfo{author}{\bibfnamefont{J.}~\bibnamefont{Evers}}, \bibnamefont{and}
  \bibinfo{author}{\bibfnamefont{C.~H.} \bibnamefont{Keitel}},
  \bibinfo{journal}{Phys.~Rev.~A} \textbf{\bibinfo{volume}{75}},
  \bibinfo{pages}{032313} (\bibinfo{year}{2007}).

\bibitem[{\citenamefont{Belavkin et~al.}(1969)\citenamefont{Belavkin,
  Zeldovich, Perelomov, and Popov}}]{Bela69}
\bibinfo{author}{\bibfnamefont{A.~A.} \bibnamefont{Belavkin}},
  \bibinfo{author}{\bibfnamefont{B.~Y.} \bibnamefont{Zeldovich}},
  \bibinfo{author}{\bibfnamefont{A.~M.} \bibnamefont{Perelomov}},
  \bibnamefont{and} \bibinfo{author}{\bibfnamefont{V.~S.} \bibnamefont{Popov}},
  \bibinfo{journal}{Sov. Phys. JETP} \textbf{\bibinfo{volume}{56}},
  \bibinfo{pages}{264} (\bibinfo{year}{1969}).

\bibitem[{\citenamefont{Lehmberg}(1970{\natexlab{a}})}]{Lehm70i}
\bibinfo{author}{\bibfnamefont{R.~H.} \bibnamefont{Lehmberg}},
  \bibinfo{journal}{Phys. Rev. A} \textbf{\bibinfo{volume}{2}},
  \bibinfo{pages}{883} (\bibinfo{year}{1970}{\natexlab{a}}).

\bibitem[{\citenamefont{Lehmberg}(1970{\natexlab{b}})}]{Lehm70ii}
\bibinfo{author}{\bibfnamefont{R.~H.} \bibnamefont{Lehmberg}},
  \bibinfo{journal}{Phys. Rev. A} \textbf{\bibinfo{volume}{2}},
  \bibinfo{pages}{889} (\bibinfo{year}{1970}{\natexlab{b}}).

\bibitem[{\citenamefont{Agarwal}(1970)}]{Arg70}
\bibinfo{author}{\bibfnamefont{G.~S.} \bibnamefont{Agarwal}},
  \bibinfo{journal}{Phys. Rev. A} \textbf{\bibinfo{volume}{2}},
  \bibinfo{pages}{2038} (\bibinfo{year}{1970}).

\bibitem[{\citenamefont{Gross and Haroche}(1982)}]{Gro}
\bibinfo{author}{\bibfnamefont{M.}~\bibnamefont{Gross}} \bibnamefont{and}
  \bibinfo{author}{\bibfnamefont{S.}~\bibnamefont{Haroche}},
  \bibinfo{journal}{Phys. Rep} \textbf{\bibinfo{volume}{93}},
  \bibinfo{pages}{301} (\bibinfo{year}{1982}).

\bibitem[{\citenamefont{Carmichael and Kim}(2000)}]{Car00}
\bibinfo{author}{\bibfnamefont{H.~J.} \bibnamefont{Carmichael}}
  \bibnamefont{and} \bibinfo{author}{\bibfnamefont{K.}~\bibnamefont{Kim}},
  \bibinfo{journal}{Opt. Commun.} \textbf{\bibinfo{volume}{179}},
  \bibinfo{pages}{417} (\bibinfo{year}{2000}).

\bibitem[{\citenamefont{Clemens et~al.}(2003)\citenamefont{Clemens, Horvath,
  Sanders, and Carmichael}}]{Clem03}
\bibinfo{author}{\bibfnamefont{J.~P.} \bibnamefont{Clemens}},
  \bibinfo{author}{\bibfnamefont{L.}~\bibnamefont{Horvath}},
  \bibinfo{author}{\bibfnamefont{B.~C.} \bibnamefont{Sanders}},
  \bibnamefont{and} \bibinfo{author}{\bibfnamefont{H.~J.}
  \bibnamefont{Carmichael}}, \bibinfo{journal}{Phys. Rev. A}
  \textbf{\bibinfo{volume}{68}}, \bibinfo{pages}{023809}
  (\bibinfo{year}{2003}).

\bibitem[{\citenamefont{Singer et~al.}(2005)\citenamefont{Singer, Stanojevic,
  Weidem\"{u}ller, and C\^{o}t\'{e}}}]{singer2005}
\bibinfo{author}{\bibfnamefont{K.}~\bibnamefont{Singer}},
  \bibinfo{author}{\bibfnamefont{J.}~\bibnamefont{Stanojevic}},
  \bibinfo{author}{\bibfnamefont{M.}~\bibnamefont{Weidem\"{u}ller}},
  \bibnamefont{and}
  \bibinfo{author}{\bibfnamefont{R.}~\bibnamefont{C\^{o}t\'{e}}},
  \bibinfo{journal}{J. Phys. B} \textbf{\bibinfo{volume}{38}},
  \bibinfo{pages}{S295} (\bibinfo{year}{2005}).

\bibitem[{\citenamefont{Bethe}(1947)}]{Bet47}
\bibinfo{author}{\bibfnamefont{H.~A.} \bibnamefont{Bethe}},
  \bibinfo{journal}{Phys. Rev.} \textbf{\bibinfo{volume}{72}},
  \bibinfo{pages}{339} (\bibinfo{year}{1947}).

\bibitem[{\citenamefont{de~Vries et~al.}(1998)\citenamefont{de~Vries, van
  Coevorden, and Lagendijk}}]{Vrie}
\bibinfo{author}{\bibfnamefont{P.}~\bibnamefont{de~Vries}},
  \bibinfo{author}{\bibfnamefont{D.~V.} \bibnamefont{van Coevorden}},
  \bibnamefont{and}
  \bibinfo{author}{\bibfnamefont{A.}~\bibnamefont{Lagendijk}},
  \bibinfo{journal}{Rev. Mod. Phys.} \textbf{\bibinfo{volume}{70}},
  \bibinfo{pages}{447} (\bibinfo{year}{1998}).

\bibitem[{\citenamefont{Coffey and Friedberg}(1978)}]{Coff78}
\bibinfo{author}{\bibfnamefont{B.}~\bibnamefont{Coffey}} \bibnamefont{and}
  \bibinfo{author}{\bibfnamefont{R.}~\bibnamefont{Friedberg}},
  \bibinfo{journal}{Phys. Rev. A} \textbf{\bibinfo{volume}{17}},
  \bibinfo{pages}{1033} (\bibinfo{year}{1978}).

\bibitem[{\citenamefont{Zanardi}(1998)}]{Zan98}
\bibinfo{author}{\bibfnamefont{P.}~\bibnamefont{Zanardi}},
  \bibinfo{journal}{Phys. Rev. A} \textbf{\bibinfo{volume}{57}},
  \bibinfo{pages}{3276} (\bibinfo{year}{1998}).

\bibitem[{\citenamefont{Sakurai}(1976)}]{Sak76}
\bibinfo{author}{\bibfnamefont{J.~J.} \bibnamefont{Sakurai}},
  \emph{\bibinfo{title}{Advanced Quantum Mechanics}}
  (\bibinfo{publisher}{Addison-Wesley}, \bibinfo{address}{Reading, Mass.},
  \bibinfo{year}{1976}).

\bibitem[{\citenamefont{Power and Zienau}(1957)}]{Pow57}
\bibinfo{author}{\bibfnamefont{E.~A.} \bibnamefont{Power}} \bibnamefont{and}
  \bibinfo{author}{\bibfnamefont{S.}~\bibnamefont{Zienau}},
  \bibinfo{journal}{Il Nuovo Cimento} \textbf{\bibinfo{volume}{6}},
  \bibinfo{pages}{7} (\bibinfo{year}{1957}).

\bibitem[{\citenamefont{Friedberg et~al.}(1973)\citenamefont{Friedberg,
  Hartmann, and Manassah}}]{Fri73}
\bibinfo{author}{\bibfnamefont{R.}~\bibnamefont{Friedberg}},
  \bibinfo{author}{\bibfnamefont{S.~R.} \bibnamefont{Hartmann}},
  \bibnamefont{and} \bibinfo{author}{\bibfnamefont{J.~T.}
  \bibnamefont{Manassah}}, \bibinfo{journal}{Phys. Reps.}
  \textbf{\bibinfo{volume}{7}}, \bibinfo{pages}{101} (\bibinfo{year}{1973}).

\bibitem[{\citenamefont{Agarwal}(1974)}]{Aga74}
\bibinfo{author}{\bibfnamefont{G.~S.} \bibnamefont{Agarwal}},
  \emph{\bibinfo{title}{Quantum Statistical Theories of Spontaneous Emission
  and their Relation to Other Approaches}}
  (\bibinfo{publisher}{Springer-Verlag}, \bibinfo{address}{Berlin, Germany},
  \bibinfo{year}{1974}).

\bibitem[{\citenamefont{Power and Zienau}(1959)}]{Pow59}
\bibinfo{author}{\bibfnamefont{E.~A.} \bibnamefont{Power}} \bibnamefont{and}
  \bibinfo{author}{\bibfnamefont{S.}~\bibnamefont{Zienau}},
  \bibinfo{journal}{Phil. Trans. Roy. Soc. A} \textbf{\bibinfo{volume}{251}},
  \bibinfo{pages}{427} (\bibinfo{year}{1959}).

\bibitem[{\citenamefont{Milonni}(1994)}]{Mil94}
\bibinfo{author}{\bibfnamefont{P.~W.} \bibnamefont{Milonni}},
  \emph{\bibinfo{title}{The Quantum Vacuum: An Introduction to Quantum
  Electrodynamics}} (\bibinfo{publisher}{Academic Press, Inc.},
  \bibinfo{address}{London, UK}, \bibinfo{year}{1994}).

\bibitem[{\citenamefont{Cohen-Tannoudji
  et~al.}(1989)\citenamefont{Cohen-Tannoudji, Dupont-Roc, and
  Grynberg}}]{Coh89}
\bibinfo{author}{\bibfnamefont{C.}~\bibnamefont{Cohen-Tannoudji}},
  \bibinfo{author}{\bibfnamefont{J.}~\bibnamefont{Dupont-Roc}},
  \bibnamefont{and} \bibinfo{author}{\bibfnamefont{G.}~\bibnamefont{Grynberg}},
  \emph{\bibinfo{title}{Photons and Atoms: Introduction to Quantum
  Electrodynamics}} (\bibinfo{publisher}{Wiley}, \bibinfo{address}{New York},
  \bibinfo{year}{1989}).

\bibitem[{\citenamefont{Davidovich}()}]{Dav75}
\bibinfo{author}{\bibfnamefont{L.}~\bibnamefont{Davidovich}},
  \bibinfo{note}{{P}h.D. Thesis, Department of Physics, University of
  Rochester, 1975}.

\end{thebibliography}
\end{document}